%% file: cbc2017.tex
\documentclass[review]{elsarticle}

\usepackage{hyperref}
\usepackage{algorithmic}
\usepackage{algorithm}
\usepackage[caption=false,font=normalsize,labelfont=sf,textfont=sf]{subfig}
\usepackage{booktabs}
\usepackage{url}

\journal{Journal of \LaTeX\ Templates}

\begin{document}

\begin{frontmatter}

\title{Estimation of Distribution Algorithm for Protein Structure Prediction}

\author{Daniel Bonetti}
\cortext[mycorrespondingauthor]{Corresponding author}
\ead{daniel.bonetti@gmail.com}

\author{Alexandre Delbem}
\author{Dorival Le\~ao}
\address{Institute of Mathematical Sciences and Computation, University of Sao Paulo, SP, Brazil}

\author{Jochen Einbeck}
\address{Department of Mathematical Sciences, Durham University, Durham, UK}

%\author[mymainaddress,mysecondaryaddress] {Daniel Bonetti}


\begin{abstract}

% metade desse abstract veio do artigo da plos
  Proteins are essential for maintaining life. For example, knowing the structure of a protein, cell regulatory mechanisms of organisms can be modeled, supporting
  the development of disease treatments or the understanding of relationships between protein structures and food attributes. However, discovering the structure
  of a protein can be a difficult and expensive task, since it is hard to explore the large search to predict even a small protein. Template-based methods
  (coarse-grained, homology, threading etc) depend on Prior Knowledge (PK) of proteins determined using other methods as X-Ray Crystallography or Nuclear
  Magnetic Resonance. On the other hand, template-free methods (full-atom and \emph{ab initio}) rely on atoms physical-chemical properties to predict protein
  structures. In comparison with other approaches, the Estimation of Distribution Algorithms (EDAs) can require significant less PK, suggesting that it could
  be adequate for proteins of low-level of PK. Finding an EDA able to handle both prediction quality and computational time is a difficult task, since they
  are strong inversely correlated. We developed an EDA specific for the \emph{ab initio} Protein Structure Prediction (PSP) problem using full-atom representation.
  We developed one univariate and two bivariate probabilistic models in order to design a proper EDA for PSP. The bivariate models make relationships between
  dihedral angles $\phi$ and $\psi$ within an amino acid. Furthermore, we compared the proposed EDA with other approaches from the literature. We noticed
  that even a relatively simple algorithm such as Random Walk can find the correct solution, but it would require a large amount of prior knowledge
  (biased prediction). On the other hand, our EDA was able to correctly predict with no prior knowledge at all, characterizing such a prediction as
  pure \emph{ab initio}.
\end{abstract}

\begin{keyword}
Estimation of Distribution Algorithm\sep Protein Structure Prediction\sep \emph{ab initio}\sep full-atom\sep van der Waals energy\sep Expectation-Maximization.
%\texttt{elsarticle.cls}\sep \LaTeX\sep Elsevier \sep template
%\MSC[2010] 00-01\sep  99-00
\end{keyword}

\end{frontmatter}

%\linenumbers

\section{Introduction}

From the discovery of a new disease until the development of its cure can take about ten years and involve costs of
five billion dollars \cite{Herper2013}. One of the main reasons for this amount of money and time needed is the
problem of finding the tertiary structure of the protein responsible for the disease. Most of the existing
methods to determine the protein structures are experimental. They attempt to look at the proteins as they
are present in nature. For instance, the X-Ray Crystallography experimental method shoots an x-ray beam
into a protein crystal. Then it creates a diffraction map of hydrogen atoms and together with information
of the primary structure of the protein, it is possible to construct the tertiary structure. The Nuclear
Magnetic Resonance needs a solution of high concentration containing the target protein. This solution
is submitted to a process that will excite the atom spins and, depending on how much the atoms will move
to a different spin state, it is possible to construct the tertiary structure of a protein. Both of
these experimental methods have disadvantages and they cannot always determine the tertiary structure of a protein \cite{Lengauer2002}.

Thus, in order to avoid the high costs and time needed by experimental methods, computational methods (\emph{in silico})
have been created. The method to determine the primary structure of proteins is well established nowadays, but as we know,
the equivalent tertiary structure is not. Thus, from the sequence of amino acids, computational methods attempt to look
into the search space in order to find feasible protein configurations. We know that proteins in nature stabilize with
the minimum energy state. Therefore, computational methods look for protein configurations that represent the minimum
energy state. It means that we are predicting the protein structure. This problem is known as Protein Structure Prediction (PSP).
There are two main computational approaches to predict proteins. One uses knowledge of proteins that were determined using other
methods, as experimental, to infer the new ones (called template-based). Other, as used in this work, is called template-free.
It does not use information about any existing structures to predict the new ones. Instead, it uses energy potentials so it
can evaluate how good a configuration is. Then, when a configuration with a lower energy value is found, we will have a
hypothesis that we found the correct structure. However, the search space of protein configurations is huge and this
task cannot be performed using exact methods \cite{Bujnicki2009}.

Our goal is to find the correct set of the dihedral angles $\phi, \psi, \chi$'s (Section~\ref{protRep}), i.e. the variables of the problem,
that will yield the protein configurations with the lowest possible energy. Thus, this can be treated as an optimization problem in which
we want to minimize the energy of a protein configuration (our fitness) changing the dihedral angles (the variables). There are different ways
that one can represent the protein configurations in the computer. One could use a coarse-grain representation of amino acid, called
HP model \cite{Dill1985}, but it cannot represent the protein for practical purposes. Therefore, we use the full-atom model
considering all atoms of the protein configurations \cite{Duan2001}.

Despite the efforts of the \emph{in silico} methods in trying to find the correct protein configurations, an appropriate
algorithm that can work properly in a pure \emph{ab initio} way is still missing, which is especially relevant when one
wants to predict a protein that has low similarity with the known structures. There are some Evolutionary Algorithms (EA)
and other optimization approaches with useful results for PSP \cite{Hardin2002, Lima2007a, Berenboym2008, Brasil2013}.
The better the optimization technique, the better will be its chances of finding proper solutions. Nevertheless,
this is not a trivial task since the optimization algorithm has to deal with the large search space of possible
3D protein structures in an efficient manner.

Thus, considering the characteristics of the PSP problem, we build an Estimation of Distribution Algorithm (EDA) \cite{Larranaga2002}
specially designed for such a problem. Basically, an EDA constructs models of the distribution of variable values from promising regions
of the search space in order to improve the optimization process. To sample better solutions, we need to have a good probabilistic model.
First, we developed a Univariate model-based Optimization algorithm (UNIO). Secondly, taking advantage of properties of the problem we
are tackling, we modeled the $[\phi,\psi]$ within the same residue as correlated, yielding a bivariate probabilistic model. We proposed
two different ways to estimate a probability distribution for each pair of variables: (i) Kernel Density Estimation model-based
Optimization (KDEO) \cite{Wand1995} and (ii) Finite Gaussian Mixtures model-based Optimization (FGMO) \cite{McLachlan2004}. That
yielded three novel different EDAs for PSP: UNIO, KDEO and FGMO.

In order to evaluate our probabilistic models, we measured three aspects: the quality of the protein configurations (RMSD,
root-mean-square deviation) \cite{Baxevanis2001}, the computational time consumption and the energy values of the predicted
structures (Section~\ref{results}). As we expected, the bivariate models (KDEO and FGMO) were able to find better solutions
in terms of the energy and RMSD. Furthermore, we made a comparison between KDEO and specifically other optimization approaches
from the literature, Random Walk (RW) \cite{Pearson1905}, Monte Carlo (MC) \cite{Metropolis1949}, Genetic Algorithm (GA) \cite{Goldberg1989}
and Differential Evolution (DE) \cite{Storn1997}. We discovered that all of these optimization techniques were able to find
the correct protein configuration, of a specific case, when there is enough previous knowledge about a promising region of
the search space. For instance, RW is known as a poor optimization technique, thus, in order to find the correct protein
configuration, we introduced a bias, reducing the search space to a neighborhood of the dihedral angles of the native protein.
On the other hand, the proposed EDA can, in general, find adequate solutions without search space reduction, that is, without bias.
Therefore, we may concluded that the proposed EDA is more adequate than other investigated approaches when predicting proteins
with low similarity to known structures.

In the next section, we present the Protein Structure Prediction problem. In Section~\ref{EDAforPSP}, we present our proposed
EDA for PSP. Then, the results are shown in Section~\ref{results}. Finally, the conclusions of this work are made in Section~\ref{conclusions}.

\section{Protein Structure Prediction problem}

Proteins are essential in almost all processes of life. They have different functions in living beings and the function of
each protein is associated with their shape. For instance, a protein responsible for the transportation of other molecules
has its shape appropriate to carry such a molecule along the organism until its destination, i.e. the hemoglobin, capable
to transport oxygen. There are several other functions of proteins as, for instance, defense, control, regulation,
breaking covalent binds etc. We say that a function of a protein is associated to its shape, also called tertiary structure \cite{Branden1999}.

When we know a protein responsible of causing a disease, we can develop treatments and inhibit that protein function.
For instance, some viruses have the capability of cutting human cells, entering in it and then making copies of themselves
spreading the disease along the rest of body. However, when we know the shape of the protein responsible for cutting the
cell we can develop medicines, the complementary structure, which bind the virus in the medicine molecule instead of the
human cell. This will probably lead to the death of the virus and reduce the effects of the virus on the organism \cite{Bujnicki2009}.

Every protein has an identification that matches with its shape. This is called primary structure (or sequence of amino acids)
and it stores all the residues chain of a protein. Nowadays, it is relatively simple to isolate the primary structure of a protein.
However, finding the tertiary structure equivalent to the primary structure is a very complicated task. This is the main reason why
the sequence database is growing fast \cite{UniProt2014} and the structures database is growing relatively slow \cite{PDB2014}.

The current methods to find the tertiary structure of proteins are expensive and require years of trial and error. The experimental
methods X-Ray Crystallography (XRC) and Nuclear Magnetic Resonance (NMR) are by far the most used methods to determine protein
structures. However, it is not always possible to use these methods, since they have disadvantages as, for instance, sometimes
it is not possible to have a protein crystal and in this case, the XRC will be no longer available. There is no need to have a
protein crystal in NMR but it only works for small proteins \cite{Baxevanis2001}.

Based on the drawbacks of the experimental methods many computer methods that try to find the structure of proteins were developed.
This is known as \emph{in silico} methods. Based on the sequence of amino acids, they look for feasible protein configurations in
the search space and try to predict the protein structure. The \emph{in silico} methods are divided into two different ways of
prediction. First, they are based on prior knowledge of proteins that were already determined by other methods, as XRC and NMR.
They are promising techniques and there are many researches about these methods. One problem of these techniques is that their
predictions can be biased toward the experimental methods. The second method is called template-free and does not make use of
knowledge of other proteins. Instead, it uses energy potentials in order to compute the energy of the protein configuration.
The energy potential is a way to describe how a protein configuration represents a protein in real life. Knowing that proteins
stabilize in a state of minimum energy, we also want to find the configurations with lowest energy. In this work, we use a
template-free approach with full-atom representation \cite{Bonneau2001}.

\subsection{Protein representation}
\label{protRep}

\begin{figure}[!htb]
	\centering
    \includegraphics[width=0.8\textwidth]{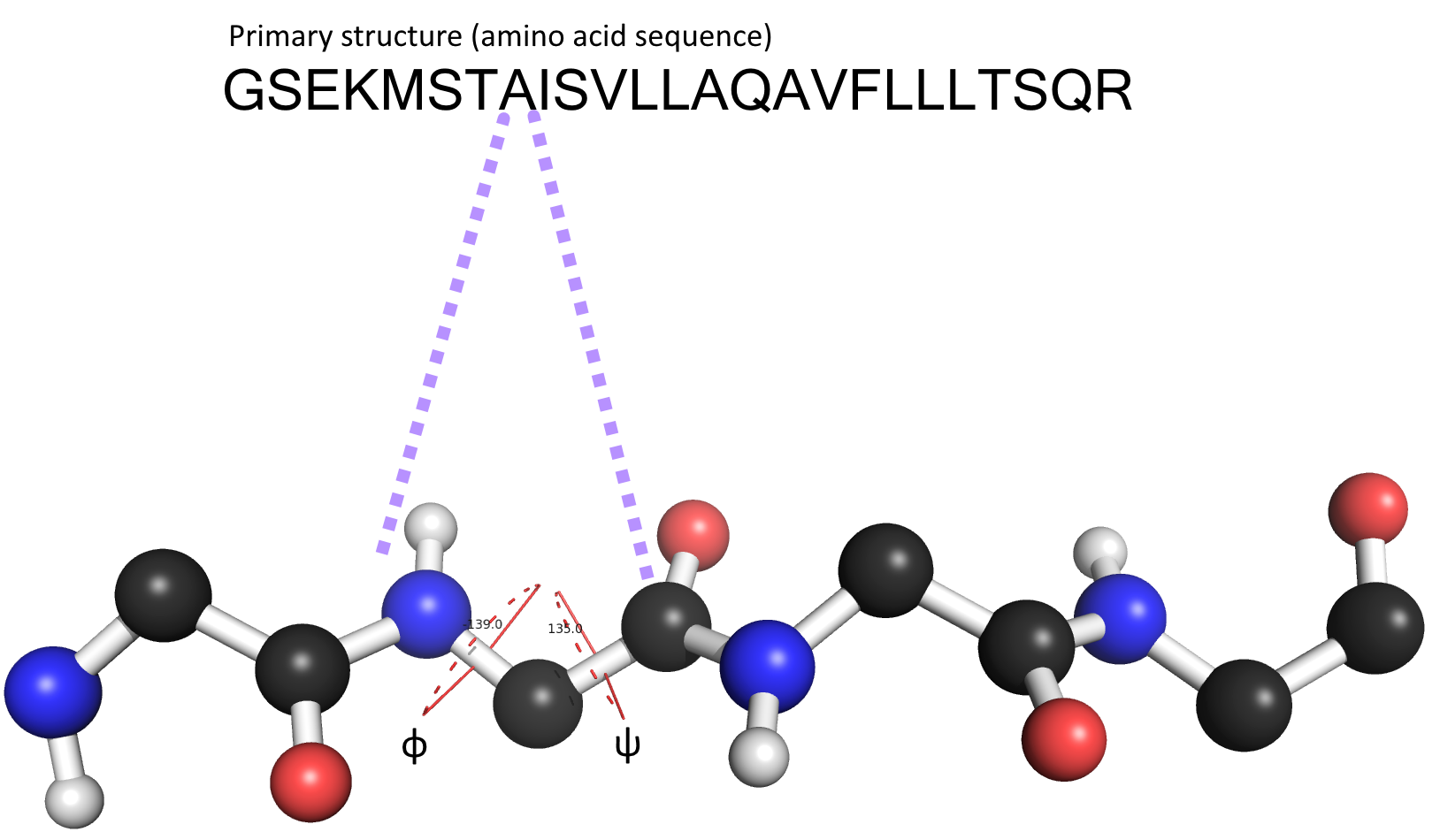}
	\caption{An example of the dihedral angles $[\phi, \psi]$ from the sequence of amino acids using a full-atom representation (the side chains are hidden).}
	\label{fig1}
\end{figure}

The protein configuration can often be represented using dihedral angles $\phi, \psi, \chi$'s. Each residue in a protein configuration
chain has its own set of $[\phi, \psi]$ angles and the number of $\chi$'s depends on the type of residue. All these angles range
from $-180$ to $180$ degrees, the search space range. The $\phi$ angle is the dihedral angle between atoms $N-C_\alpha$ within the
same residue and the $\psi$ is related to the $C_\alpha-C$ angle. The set of all dihedral angles $\phi$'s, $\psi$'s, $\chi$'s from all
residues of a protein configuration has all information needed to evaluate the protein. Thus, the variables of the problem can
be understood as a vector of dihedral angles ranging from $[-180; 180]$ (Figure~\ref{fig1}).

In order to compute the fitness of a protein configuration, we first need to convert the dihedral angles into Cartesian coordinates.
Then, knowing the Cartesian coordinates of all atoms and their types it is possible to compute the energy (the fitness).

\subsection{The fitness function}

Several potential energies contribute to the protein energy. There are bonded and non-bonded potential energies and each of them has a
specific contribution to the molecule stabilization. We know that non-bonded potential energies have the largest contribution to the
molecule energy. In our algorithm, called ProtPred \cite{Lima2007a}, we have five bonded potential energies (bond stretching, angle bending,
Urey-Bradley, improper dihedral and torsional angle) and four non-bonded potential energies (van der Waals, electrostatic, solvation and hydrogen bonds).
Despite having all these potential energies implemented in our algorithm, we decide to use only van der Waals energy in this paper, since it has the
highest contribution to molecule energy. Besides, it makes it easier to understand the results so we can keep the focus on the evolutionary process.
Thus, the fitness function is determined only by the van der Waals energy.

The van der Waals energy models the attraction and repulsion among atoms. In general, the Lennard-Jones potential (also known as Lennard-Jones $12-6$)
is used to compute the van der Waals energy of a protein configuration. The van der Waals energy changes according to the distance and the atom's
type (nitrogen, carbon, oxygen, hydrogen etc). Equation~\ref{lj} describes the relative distance between two atoms $i$ and $j$, given the
Euclidian distance between them ($d_{i,j}$) and the van der Waals radii constant $R$ of atoms $i$ and $j$:

\begin{equation}
v_{ij} = \frac{d_{i,j}}{R_{i}+R_{j}}
\label{lj}
.
\end{equation}

\begin{figure}[!htb]
	\centering
    \includegraphics[width=.75\textwidth]{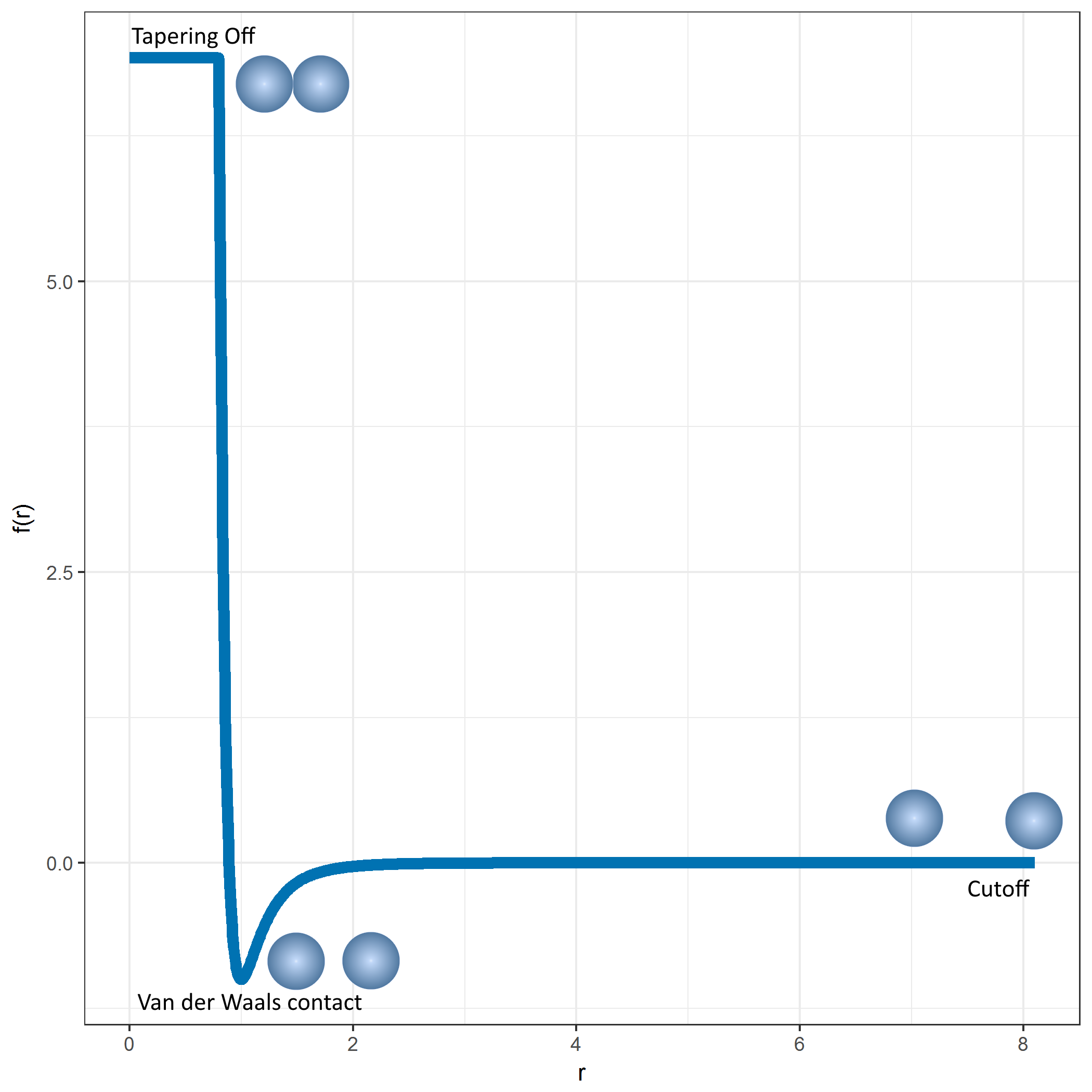}
    \caption{The modified van der Waals function we used, highlighting the tapering-off (when atoms are to close),
      the van der Waals contact (ideal distance) and the cutoff (they are too far).}
	\label{fig2}
\end{figure}

The van der Waals energy is very repulsive at short distances since the electron cloud between atoms starts to overlap. At this distance,
the energy rapidly increases and tends to infinity. The equilibrium point, known as van der Waals contact, happens when the Euclidian
distance between the atom pair is neither too far nor too close. This is the point of minimum energy of an atom pair. If the atom pair
is too far from each other they will not have any type of interaction and the energy will tend to zero. We used a cutoff of 8~\AA~in
order to avoid unnecessary computation, and to avoid dealing with large numbers we also set a tapering-off (see Figure~\ref{fig2}).
In case that $v_{ij}$ is smaller than 0.8 we assume a constant $C$ \cite{Cui1998}. The Lennard-Jones potential used in our EDA for PSP is described by:

\begin{equation}
 f_{LJ}(v_{ij}) = \left\{
\begin{array}{rc}
   Av_{ij}^{-12}-Bv_{ij}^{-6} \mbox{ if } v_{ij}>0.8, \\
 C \mbox{ if } v_{ij} \leq 0.8,
\end{array} \right.
\label{lj2}
\end{equation}
where $A$ and $B$ are constants experimentally determined based on characteristics of the environment, and $C$ is given by $Av_{ij}^{-12}-Bv_{ij}^{-6}$ with $v_{ij}=0.8$.

The van der Waals energy is the sum of the Lennard-Jones potentials between all atom pairs in a molecule. There
are $\frac{n^2 - n}{2}$ interactions, where $n$ is the number of atoms, leading to:

\begin{equation}
E_{vdw} = \sum_{i=1}^{n-1}\sum_{j=i+1}^{n}f_{LJ} \left( v_{ij} \right).
\label{vdw}
\end{equation}

Thus, in order to have the minimum global energy of a molecule it is necessary to find a compromise between the
partial energies among atoms to get the least van der Waals energy. To find good van der Waals energy values we need
to change the dihedral angles of our protein configuration. Changing the dihedral angles will lead to a change in Cartesian
coordinates of atoms in the molecule as well. Thus, there is a set of dihedral angles that will imply the best positioning of atoms, yielding a global minimum in van der Waals energy.

\section{Estimation of Distribution Algorithm for Protein Structure Prediction}
\label{EDAforPSP}

In the previous section, we concretized the problem we want to tackle. In this section, we will describe how can we find promising
solutions using EDAs, i.e. how can the EDA find a good set of dihedral angles that will express a low energy molecule.

The EDAs are a relatively new class of the EA. They are optimization techniques that use probabilistic models from a promising set
of solutions in order to sample the offspring. In some cases they also have the capability of accounting for correlations between
variables. In the literature, the EDAs for binary and discrete variables are well described, since the probabilistic models and
variable relationship are relatively easy to understand \cite{Pelikan1999}. In fact, using a simplified representation of proteins,
\cite{Santana2008} showed that it's possible to achieve relevant results with an EDA for the PSP problem.

However, dealing with dihedral angles (in the continuous search space) is relatively more difficult since it is not possible to map
the combination of all variable values. Besides, the probabilistic model should be able to deal with multimodality, since the distributions
of dihedral angles in the PSP problem are non-parametric. We also want something that can handle the variable relationship between [$\phi, \psi$]
within the same residues, so it has to be bivariate.

There are some real-valued EDAs in the literature as well. However, they would not be totally appropriate for the PSP problem.
For example, the Univariate Marginal Distribution Algorithm (UMDA$_c$) \cite{Larranaga1999} works in the continuous search space
but cannot deal with multimodality neither variable relationship. The Bivariate Marginal Distribution Algorithm (BMDA) could
model the variable relationship but not the multimodality aspect \cite{Pelikan1999a}. Since then, EDAs able to tackle the
multimodality and the variable relationship aspects were developed, as the case of the EGNA \cite{Larranaga1999},
IDEA \cite{Bosman2000a, Bosman2001} and PBIL \cite{Gallagher1999}. The real-valued Bayesian Optimization Algorithm (rBOA) \cite{Ahn2004}
can also handle both multimodality and variable relationship. It was shown that rBOA can outperform the mIDEA \cite{Hauschild2011}
in several benchmark problems. However, the evidence that rBOA is able to predict protein structures using the
full-atom representation has not been provided in the literature yet.

Knowing that the dihedral angles $\phi$ and $\psi$ within the same residue have a strong relationship, we designed an EDA
in which there is no need to learn the variable relationship: we treat the dihedral angles [$\phi, \psi$] of each residue
as correlated variables. Besides, the statistical mechanisms used in previous EDAs as, for instance, the normal kernels
and the mixtures of Gaussian distribution were used to build our new EDA for PSP full-atom. We already showed in a previous
work that EDAs can successfully be applied in PSP using full-atom and \emph{ab initio} modeling \cite{Bonetti2014a}. In this work,
we show three different approaches that were developed. The detailed information about each of them is described in the next sections.

\subsection{Univariate model-based Optimization}

The Univariate model-based Optimization (UNIO) is the simplest algorithm. It does not deal with variable relationships. Instead,
it deals with multimodality in an efficient way. From the promising individuals (selected) the process of a one-dimensional kernel
density estimation (KDE) is simulated for each variable involved in the problem. However, creating a kernel distribution for each
problem variable would require high computational costs since it is necessary to iterate between all observations (the selected size)
per variable.

KDE is based on the sums of the difference between the point of interest $x$ and all observations within a data set ($x_1, x_2,..,x_n)$ over a
bandwidth value $h$. In order to sample new values from our kernel distribution we first need to build a Probability Density Function (PDF).
This can be done by calling the KDE function as many times as is required to fill the range of the $x$ values. That would require high computing
time per generation. For instance, consider the selected size (of the EDA) is $n=500$, and the number of points needed to build the PDF is 400,
then for 100 dihedral angles (100-dimensional problem), we would need to call the KDE function at least $500 \times 400 \times 100 = 20,000,000$
at each generation of the EDA.

Thus, to keep the algorithm efficient, we extended the idea of the KDE in a different manner. Instead of creating the PDF from the kernels,
we simply take a random observation from $x$ and add a perturbation to it with the distribution $N(0,1)$.

Consider the selected set $S^i_j$ where $i=1 \ldots n$ and $j=1\ldots m$, being $n$ the size of selected individuals  and $m$ the number
of variables of the problem. We want to fill the offspring $O^k_j$ where $k=1\ldots o$ and $o$ is offspring size. For a random point
drawn from the selected $S^i$ for a variable $j$, we add some perturbation to it and put this new value into $O^k_j$ for $k=j=1$.
Then we repeat this process for $j=1\ldots m$ (for all variables of the first individual of the offspring) and then for all the
remaining individuals ($k=1\ldots o$). The Algorithm~\ref{algUni} shows how this technique works.

\begin{algorithm}[!htb]
\caption{Random Points - It creates the offspring from selected.}
\label{algUni}
\begin{algorithmic}
\REQUIRE Selected individuals $S$, Size of selected $n$, Size of the Offspring $o$, number of problem variables $m$
\ENSURE Offspring $O$
\FOR{$j=1$ to $m$}
\STATE $u \leftarrow$ Sample $o$ values from Uniform Discrete Distribution ranging from $[1, n]$

\STATE $O_j \leftarrow S_j^u + N(0, 1)$
\ENDFOR
\end{algorithmic}
\end{algorithm}

We carried out an experiment to find out whether this kernel simplification is promising.  We noticed that the simplification of KDE,
indeed, does not have the same accuracy of the KDE, but it is much faster. So we are compromising a little of accuracy with a lot of
reduction in computational cost.

\begin{figure*}[!htb]

    \centering
      \subfloat[][]{%
        \label{natives1}%
        \includegraphics[width=.6\textwidth]{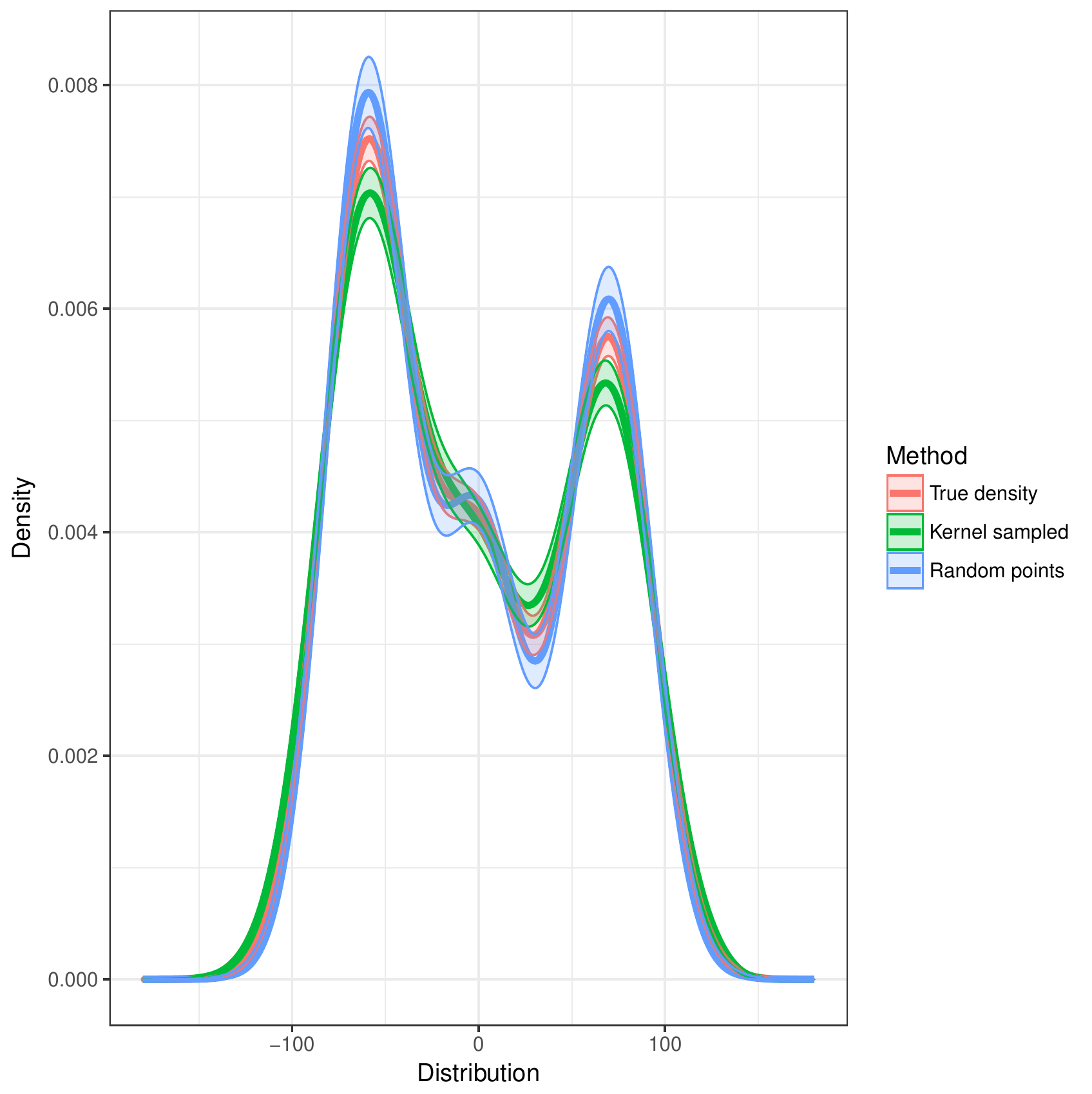}
   }

    \subfloat[][]{%
        \label{ppDensity1}%
        \includegraphics[width=.6\textwidth]{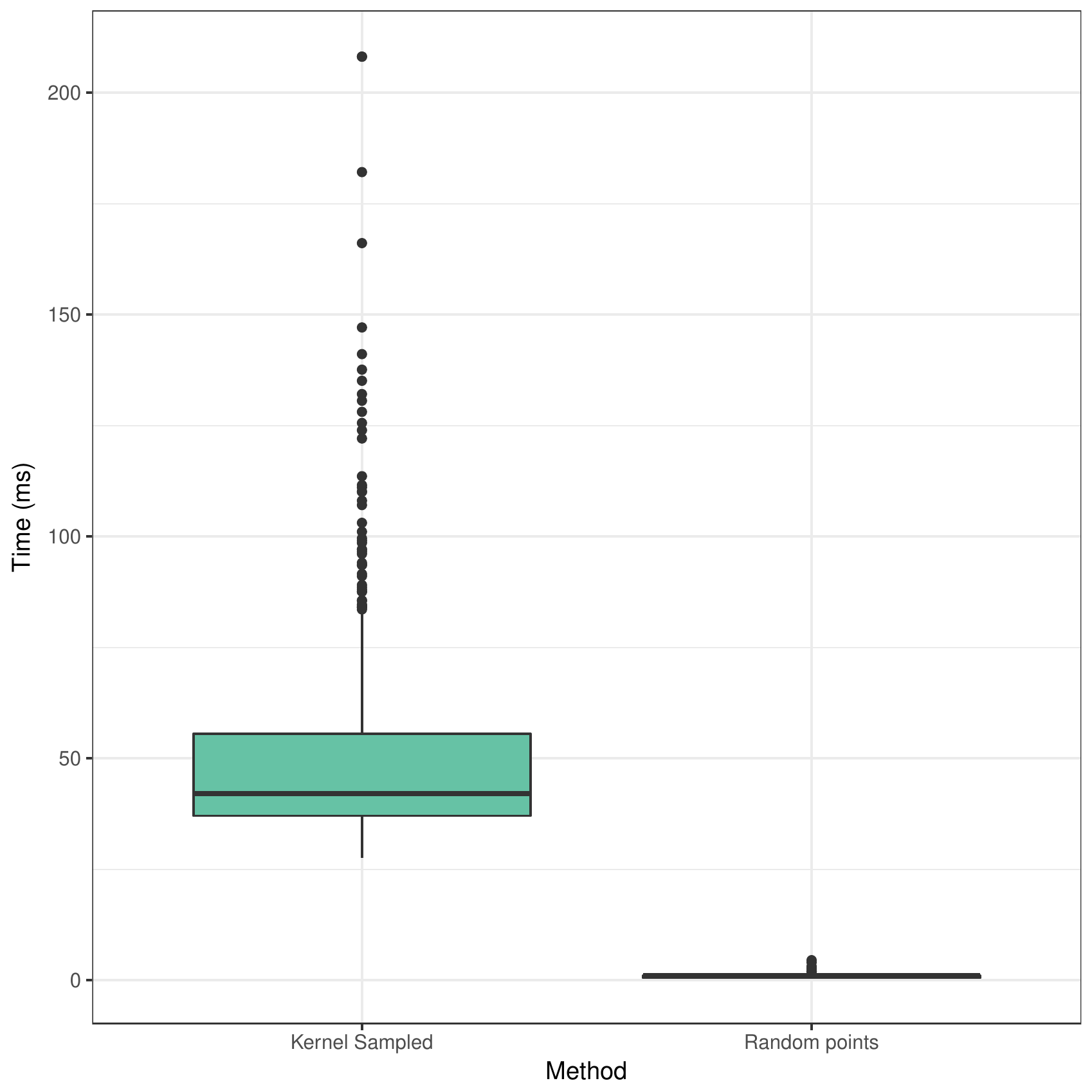}
}
    \caption[EDA.]{Comparison of simulating data with KDE and Random Points:
    \subref{natives} a comparison with the true density of the data and
    \subref{ppDensity} a time comparison.}%
    \label{unikde}%
\end{figure*}

Figure~\ref{unikde} shows an example of the comparison between our proposed simplification (Random points) and the traditional KDE.
From a sample data set with three modes, we simulated new values using KDE and our proposed simplification. We repeated this procedure
30 times. The red line represents the true density. Density plots of the simulated values with KDE and our proposed simplification
are shown as well as the smoothed standard deviations.

\subsection{Kernel Density Estimation model-based Optimization}

Considering  that all variables can interact with each other we could create an $m$--dimensional KDE. Nevertheless, that would
be very difficult to handle. Then, we decided to use the two-dimensional Kernel Density Estimation, yielding the KDEO (Kernel
Density Estimation model-based Optimization). It correlates the dihedrals angles $[\phi, \psi]$ of amino acids within the same
residue. We know that these two variables are strongly correlated since rotations in $\phi$ generally produce stereochemical
constraints for its closest neighbor dihedral angle $\psi$. Moreover, an implicit correlation is made with $[\phi, \psi]$
within the same residue every time one looks at the Ramachandran plot \cite{Ramachandran1963}, since the plot itself
shows a density map between angles $\phi$ and $\psi$.

In KDEO, the $\psi$ values are generated conditional on $\phi$. Firstly, the KDE is created for the $\phi$ and a new value $\phi'$ from
its distribution is sampled. Then, a two-dimensional KDE map of $[\phi, \psi]$ is created. The closest value to $\phi'$ in the
two-dimensional KDE (in $x$ axis direction) is taken to be the conditional KDE. Finally, a new $\psi'$ value is sampled from the
two-dimensional KDE (in $y$ axis direction).

For the two--dimensional case, kernel density estimates are obtained via Equation~\ref{kde2d1}:

\begin{equation}
\hat{f_h}(x_1, x_2) = \frac{1}{n} \sum_{i=1}^{n} \frac{1}{h_1h_2} K\left( \frac{x_1-\phi_{i}}{h_1} \right) K \left(\frac{x_2-\psi_{i}}{h_2} \right),
\label{kde2d1}
\end{equation}
where $K(.)$ is the kernel. In this work, we used the normal kernel

\begin{equation}
K(x) = (2\pi)^{-1}e^{-0.5x^2}
\label{kdegauss}
\end{equation}
and the bandwidth $\hat{h}$ is calculated using Equation~\ref{bandH}:

\begin{equation}
 \hat{h} = \left\{
\begin{array}{rc}
   4 \cdot 1.06 \cdot \hat{\sigma} \cdot n^{-1/5} \mbox{ if }  \hat{\sigma} < t, \\
   4 \cdot 1.06 \cdot t \cdot n^{-1/5} \mbox{ if } \hat{\sigma} \geq t,
\end{array} \right.
\label{bandH}
\end{equation}
where $4$ is a multiplicative factor \cite{Silverman1986, Ripley1994}  and $t$ equals

\begin{equation}
t = \frac{Q3 - Q1} {1.34},
\label{eqr}
\end{equation}
where $Q3$ and $Q1$ are the third and first quartiles, respectively.

All of these processes require many computational resources. In order to speed up the evolutionary process we first check
if the $[\phi, \psi]$ has a normal distribution using the Anderson-Darling test. If the p-value is large than 5\%, then
the KDEO is bypassed and a bivariate normal distribution is used instead. Otherwise, the KDEO is used. This strategy is
interesting because at the beginning of the evolutionary process we can have many modes. However, as the evolutionary process
starts to converge the kernel is not necessary anymore for all pairs $[\phi, \psi]$ (see Algorithm~\ref{algkde2d}).

\begin{algorithm}
\caption{Two-dimensional Kernel Sampling - It creates the Offspring from selected}
\label{algkde2d}
\begin{algorithmic}
\REQUIRE Selected individuals $S$, the number of protein residues $r$, size of the new data $o$
\ENSURE Offspring $O$
\FOR{$i=1$ to $r$ }
    \STATE $\phi \leftarrow$ Get $\phi$ from residue $i$ from $S$
    \STATE $\psi \leftarrow$ Get $\psi$ from residue $i$ from $S$
    \IF{p-value from Anderson-Darling test of $[\phi, \psi] < 0.05$}

        \STATE $\phi' \leftarrow$ Random Points ($\phi, o$)
        \STATE $P \leftarrow$ 2D Kernel Density Map ($\phi, \psi$)
        \STATE $\psi' \leftarrow$ Sample from PDF $P$ conditional on~$\phi'$
    \ELSE
        \STATE $\phi', \psi' \leftarrow$ Sample $o$ values from two-dimensional Gaussian $N([\mu_\phi, \mu_\psi], \Sigma_{\phi\psi})$
    \ENDIF
    \STATE $O^i \leftarrow $ $\phi', \psi'$
\ENDFOR
\end{algorithmic}
\end{algorithm}

\begin{figure*}%

    \centering
      \subfloat[][]{%
        \label{kde2dA}%
        \includegraphics[width=.5\textwidth]{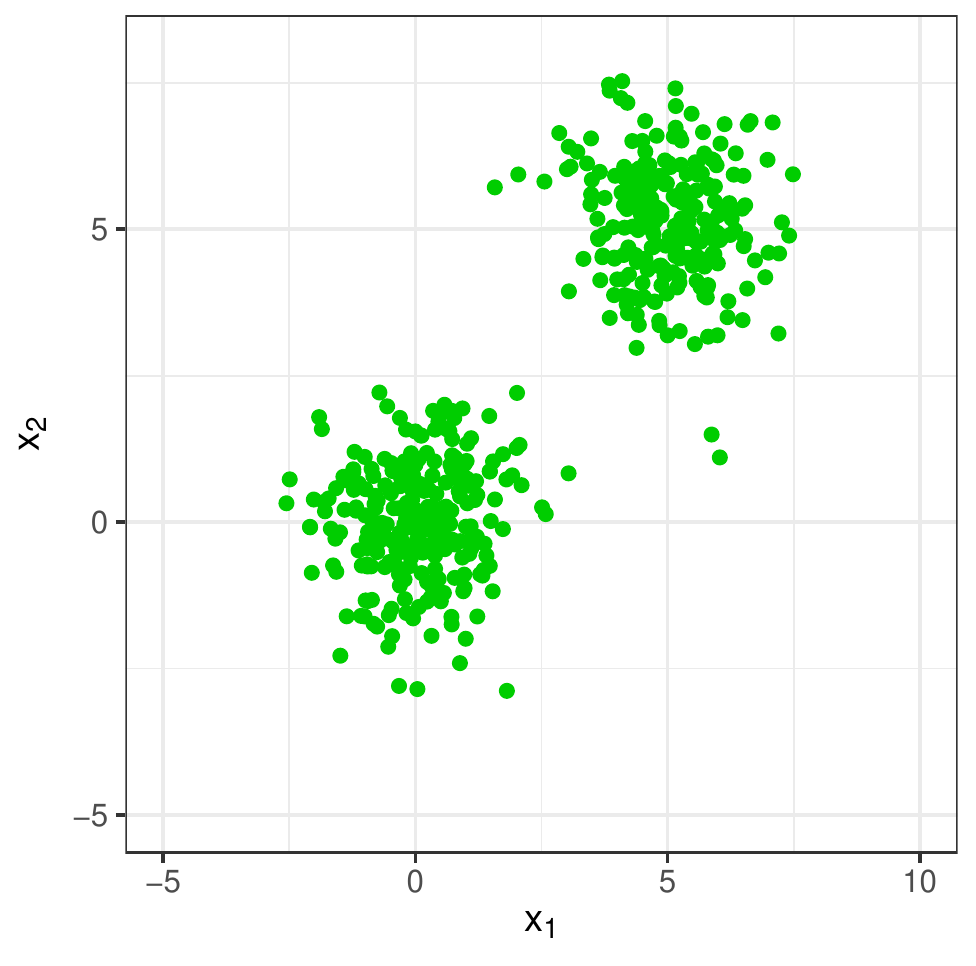}
 }
    \subfloat[][]{%
        \label{kde2dB}%
        \includegraphics[width=.4\textwidth]{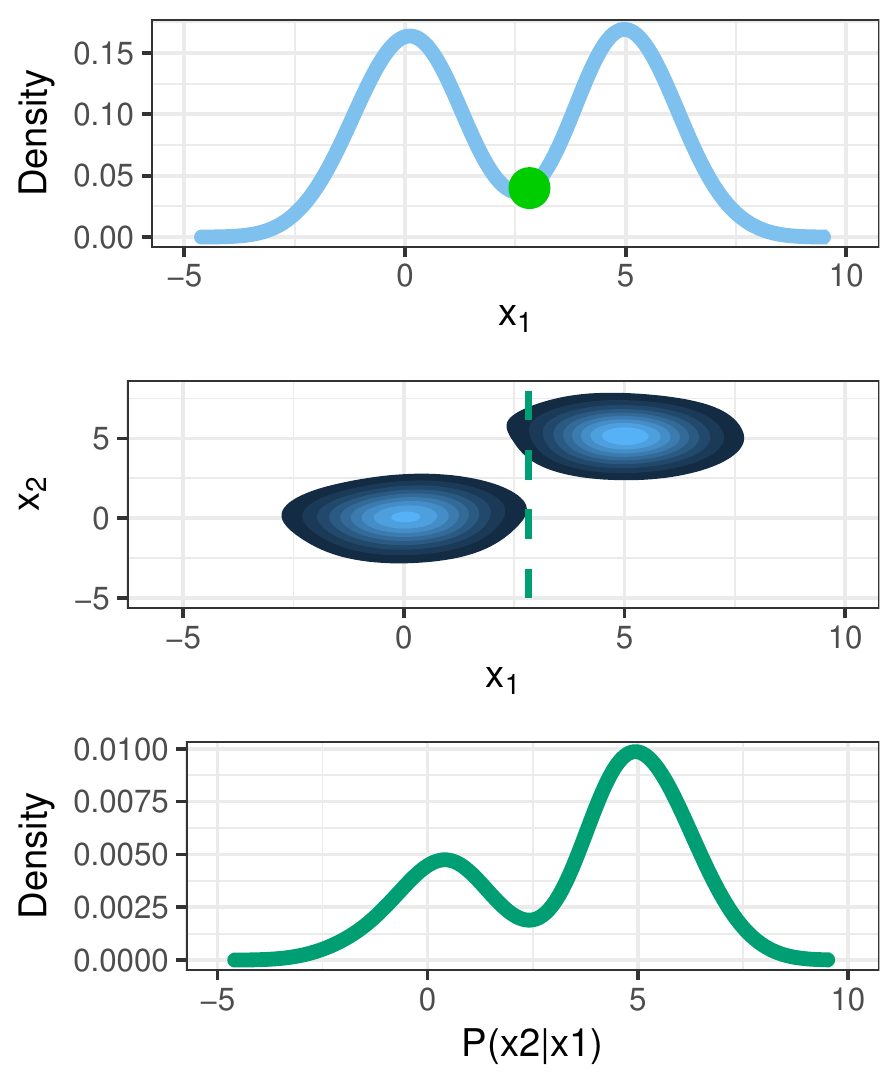}
 }

    \caption[EDA.]{Two-dimensional KDE example:
    \subref{kde2dA} The data $x_1, x_2$,
    \subref{kde2dB} the univariate distribution of $x_1$ (top), the kernel map created for $x_1,x_2$ (middle)
    and the distribution of $x_2$ given the value of $x_1$ (bottom).}%
    \label{kde2d}%
\end{figure*}

For instance, similarly to $[\phi, \psi]$, let $x_1$ and $x_2$ be two vectors of data distributed as shown in Figure~\ref{kde2dA}.
First, a new vector $x_1'$ is sampled from the independent distribution of $x_1$. Then, the two-dimensional KDE map is created
for the pair $[x_1,x_2]$. For each new point of $x_1'$ new points of $x_2'$ are sampled conditional to the previous value of $x_1'$.
Looking at the Figure~\ref{kde2dB}, let's assume that $x_1' = 3.5$ (top). As the two-dimensional KDE map must be represented using
discrete variables for $x_1$ and $x_2$, we need to pick the closest point where $x_1=3.5$ (middle) and use this point as the
conditional distribution (bottom). Finally, for each sampled $x_1'$ we sample new $x_2'$ values\footnote{The complete animation
  can be accessed at: \\ \url{http://lcrserver.icmc.usp.br/~daniel/ani/kde2d/}}.

\subsection{Finite Gaussian Mixtures model-based Optimization}

As a KDEO alternative, we develop an EDA called Finite Gaussian Mixtures model-based Optimization (FGMO). Despite having the
implementation of the $m$--dimensional FGMO we used only the two-dimensional in this paper. The FGMO combines mixtures of Gaussian
distributions in order to estimate the density of data non--parametrically. Each mixture component $k$ (cluster) has its own
mean $\mu_k$, variance matrix $\Sigma_k$ and the mixture weight $\pi_k$. From a given number of mixtures $K$, we want to know
their set of parameters $\theta = [\mu, \Sigma, \pi]$ for all mixture components.
 %\textcolor{green}{(see further comments to this section in my e--mail)}
The Expectation-Maximization (EM) is often used to estimate the parameters of Gaussian mixtures \cite{Moon1996}. In order to
get the estimated $\hat{\theta}$, the EM algorithm takes a starting setting of parameters $\hat{\theta}_0$ and iterates between
E-Step and M-Step. The algorithm converges when the log-likelihood between two iterations is less than a defined value (1.5 in
our experiments). The E-Step updates a probability matrix $w_{j,k}$ via:

\begin{equation}
    w_{j,k} =
    \frac{\hat{\pi}_k           f(x_j, \hat{\mu}_k, \hat{\Sigma}_k) }
    {\sum_{l=1}^{K} \hat{\pi}_l f(x_j, \hat{\mu}_l, \hat{\Sigma}_l) },
    \label{wjk1}
\end{equation}
where $f(x, \hat{\mu}, \hat{\Sigma}) $ can be defined as %Equation~\ref{wjk2}:

\begin{equation}
    f(x, \hat{\mu}, \hat{\Sigma})
    = \frac{1}{(2\pi)^{p/2} |\hat{\Sigma}|^{1/2} } \exp \left\{ - \frac{1}{2} (x-\hat{\mu})^T \hat{\Sigma}^{-1} (x-\hat{\mu})  \right\},
    \label{pdf1}
\end{equation}

% Old Equation 13
%\begin{equation}
%f(x) = \frac{1}{(2\pi)^{p/2} |\Sigma|^{1/2} } \exp \left\{ - \frac{1}{2} (x-\mu)^T \Sigma^{-1} (x-\mu)  \right\}
%\label{pdf1}
%\end{equation}

% Old Equation 9
%\begin{equation}
%    f(x, \hat{\mu}, \hat{\Sigma})
%    = |\hat{\Sigma}|^{-1/2} \exp \left\{ - \frac{1}{2} (x-\hat{\mu})^T \hat{\Sigma}^{-1} (x-\hat{\mu})  \right\},
%    \label{wjk2}
%\end{equation}
\noindent and the M-Step updates the parameters via

\begin{equation}
\hat{\pi}_k = \frac{1}{n} \sum_{i=1}^{n} w_{ik},
\label{kdegauss3}
\end{equation}

\begin{equation}
\hat{\mu}_k = \frac{ \sum_{i=1}^{n} w_{ik}x_i } { \sum_{i=1}^{n} w_{ik} },
\label{kdegauss4}
\end{equation}

\begin{equation}
\hat{\Sigma}_k = \frac{1}{\sum_{i=1}^nw_{ik}} \sum_{i=1}^{n} w_{ik}(x_i-\hat{\mu}_k)(x_i-\hat{\mu}_k)^T.
\label{kdegauss5}
\end{equation}

For each pair $[\phi, \psi]$ from the selected set we run a two-dimensional EM for a given number of mixture components $K$,
yielding the estimates $\hat{\theta}$. From $\hat{\theta}$ we randomly select a mixture component using a uniform distribution
with weights $\pi_k$ and sample the new $[\phi', \psi']$ values at once, using their parameters (see Algorithm~\ref{algFgmSim}).
The $\hat{\theta}$ is then used to sample all the different offspring individuals. Then, for the $[\phi, \psi]$ of the next
residue we estimate a new $\hat{\theta}$ and use these new parameters to sample the values for this next residue. This process
continues until it reaches the last residue of the molecule.

For instance, consider a protein with 5 residues. The selected individuals will have 5~$[\phi, \psi]$
pairs ($[\phi^i_1, \psi^i_1]; \ldots; [\phi^i_5, \psi^i_5]$), where $i=1 \ldots n$ being $i$ the individual
index in the selected set and $n$ is the total number of selected individuals. For the pair $[\phi^i_1, \psi^i_1]$ we perform
a complete EM algorithm and get $\hat{\theta}_1$. It contains the $\mu_k$, $\pi_k$ and $\Sigma_k$ for the mixtures $k$.
Then, the pair $[\phi{^i_1}', \psi{^i_1}']$ is generated by sampling values from $\hat{\theta}_1$. Next, $\hat{\theta}_2$ is
estimated using $[\phi^i_2, \psi^i_2]$ and then a bivariate mixture with new parameter $[\phi{^i_2}', \psi{^i_2}']$ is simulated.
This continues until the residue number 5. It is important to notice that in this case, there are five EM algorithms running at
each generation. This can have high computational costs since the EM is an iterative process. In order to speed up the algorithm,
we replaced the general PDF (Equation~\ref{pdf1})
%\begin{equation}
%f(x) = \frac{1}{(2\pi)^{p/2} |\Sigma|^{1/2} } \exp \left\{ - \frac{1}{2} (x-\mu)^T \Sigma^{-1} (x-\mu)  \right\}
%\label{pdf1}
%\end{equation}
by a specific expression for the two-dimensional case, as shown in Equation~\ref{pdf2}:

\begin{equation}
f(x_1, x_2) = \frac{1}{2 \pi \sigma_1 \sigma_2 \sqrt{1-\rho^2} } \exp \left\{ - \frac{z}{2(1-\rho^2)} \right\},
\label{pdf2}
\end{equation}

\[
z \equiv \frac{ (x_1 - \mu_1)^2 } {\sigma_1^2 } + \frac{ 2\rho(x_1 - \mu_2)(x_1 - \mu_2) } {\sigma_1 \sigma_2} + \frac{ (x_2 - \mu_2)^2 } {\sigma_2^2 }.
\label{pdfZ}
\]

Besides, before computing the FGMO we check whether the standard deviation of both $\phi$ and $\psi$ is small (say 0.01).
If true, then we bypass the EM algorithm and use a bivariate normal Gaussian instead (see Algorithm~\ref{algFgm}).

During the EM iterations, we also needed to treat special cases to avoid division by zero in Equation~\ref{wjk1}, that may
happen when there is a very far outlier with a very small standard-deviation in the data set. In this case, we stop the EM
iterations and use the last valid $\hat{\theta}$.

\begin{algorithm}
\caption{Two-dimensional Finite Gaussian Mixtures based-model Optimization - Sample the offspring using FGMO}
\label{algFgm}
\begin{algorithmic}
\REQUIRE Selected individuals $S$, the number of protein residues $r$, size of the offspring $o$, mixtures components $K$
\ENSURE Offspring $O$
\FOR{$i=1$ to $r$ }
    \STATE $\phi \leftarrow$ Get $\phi$ from residue $i$ from $S$
    \STATE $\psi \leftarrow$ Get $\psi$ from residue $i$ from $S$
    \IF{Standard deviation of $\phi > 0.01$ and Standard deviation of $\psi > 0.01$}
        \STATE $\hat{\theta} \leftarrow$ Expectation-Maximization ($\phi; \psi, K$)
        \STATE $[\phi';\psi'] \leftarrow$ Sample $o$ individuals from fitted model $\hat{\theta}$
    \ELSE
        \STATE $\phi', \psi' \leftarrow$ Sample $o$ values from two-dimensional Gaussian $N([\mu_\phi, \mu_\psi], \Sigma_{\phi\psi})$
    \ENDIF
    \STATE $O^i \leftarrow [\phi';\psi']$
\ENDFOR
\end{algorithmic}
\end{algorithm}

\begin{algorithm}
\caption{Two-dimensional Finite Gaussian Mixtures - Simulating}
\label{algFgmSim}
\begin{algorithmic}
\REQUIRE The fitted model $\hat{\theta}$ and the size of the new data $n$
\ENSURE Simulated $s$
\STATE $x \leftarrow$ Sample from distribution $U(0,1)$
\STATE $c \leftarrow$ Cumulative sum of $\hat{\theta}_\pi$
\STATE $k \leftarrow 1$
    \WHILE{$x < c_k$}
        \STATE $k \leftarrow k + 1$
    \ENDWHILE
\STATE $s \leftarrow $ Sample from $N(\hat{\theta}_{\mu_k}, \hat{\theta}_{\sigma_k})$
\end{algorithmic}
\end{algorithm}

\begin{figure*}%

    \centering
      \subfloat[][]{%
        \label{figFgmA}%
        \includegraphics[width=.3\textwidth]{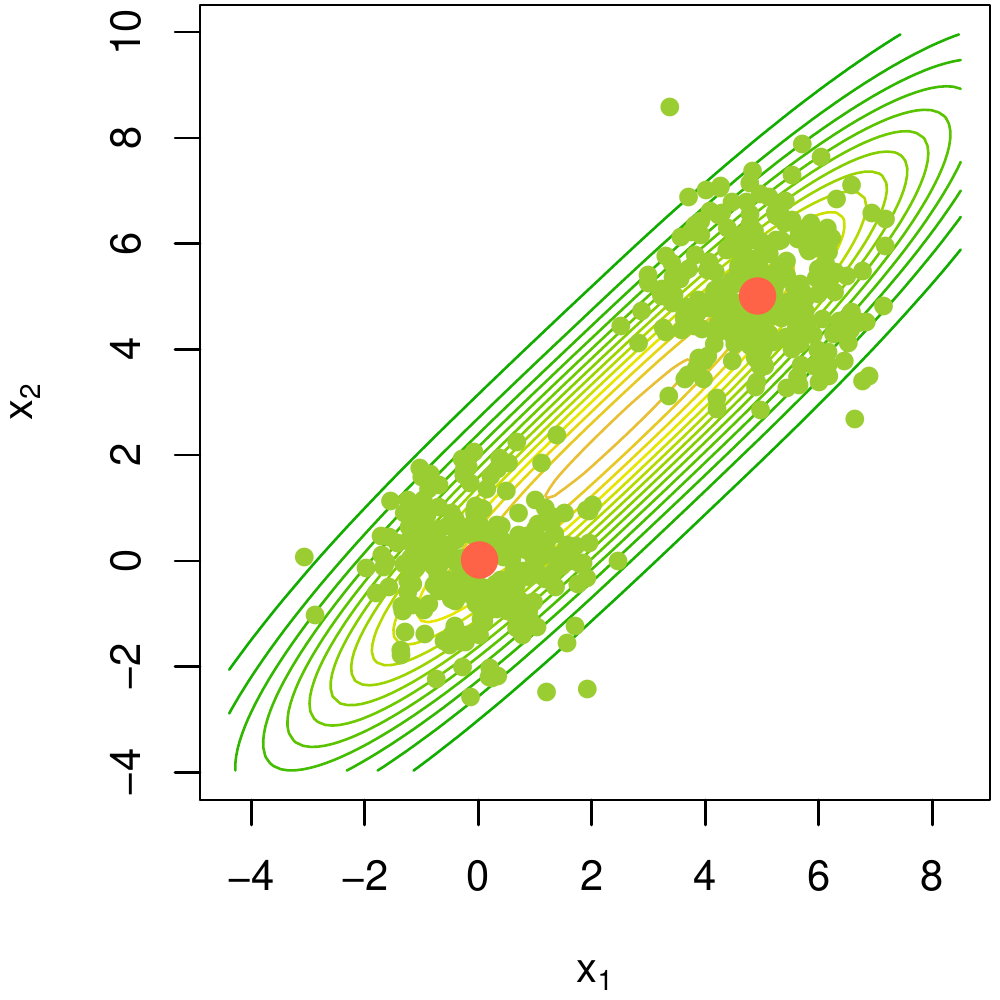}
 }
    \subfloat[][]{%
        \label{figFgmB}%
        \includegraphics[width=.3\textwidth]{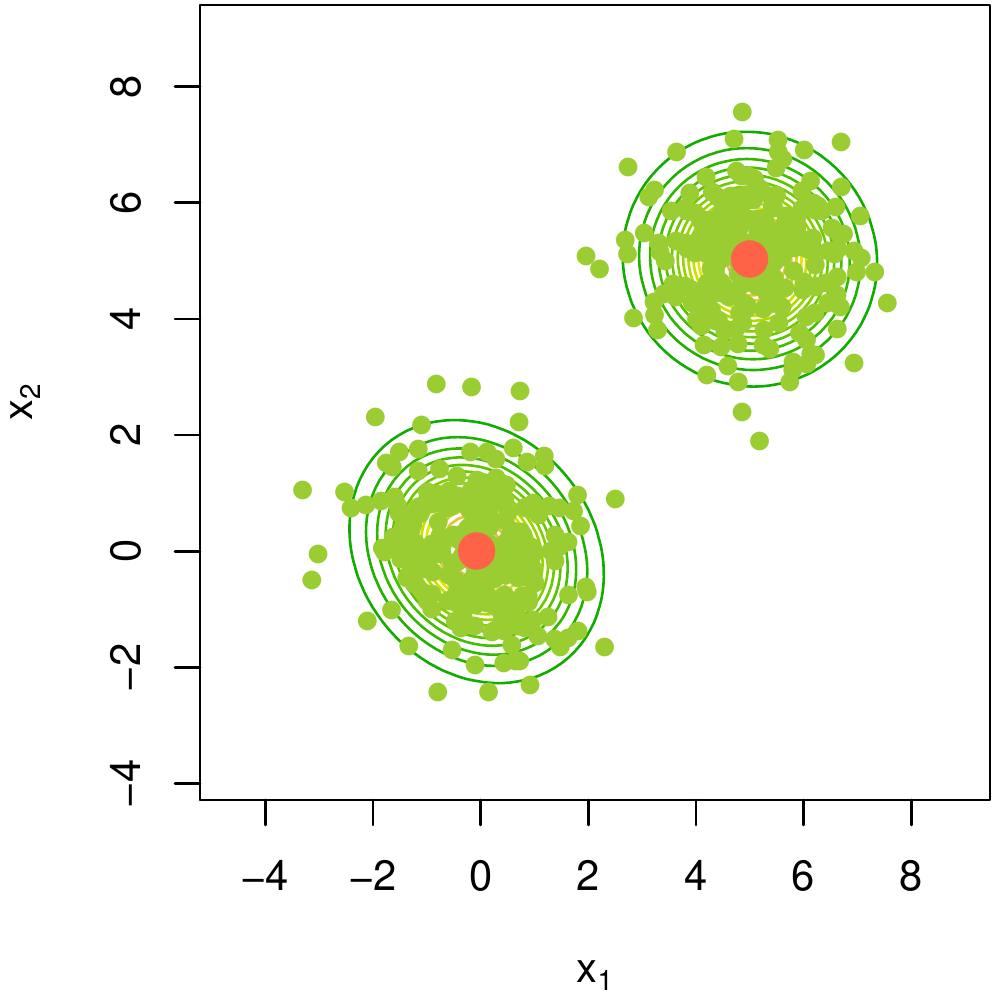}
 }
    \subfloat[][]{%
        \label{figFgmC}%
        \includegraphics[width=.3\textwidth]{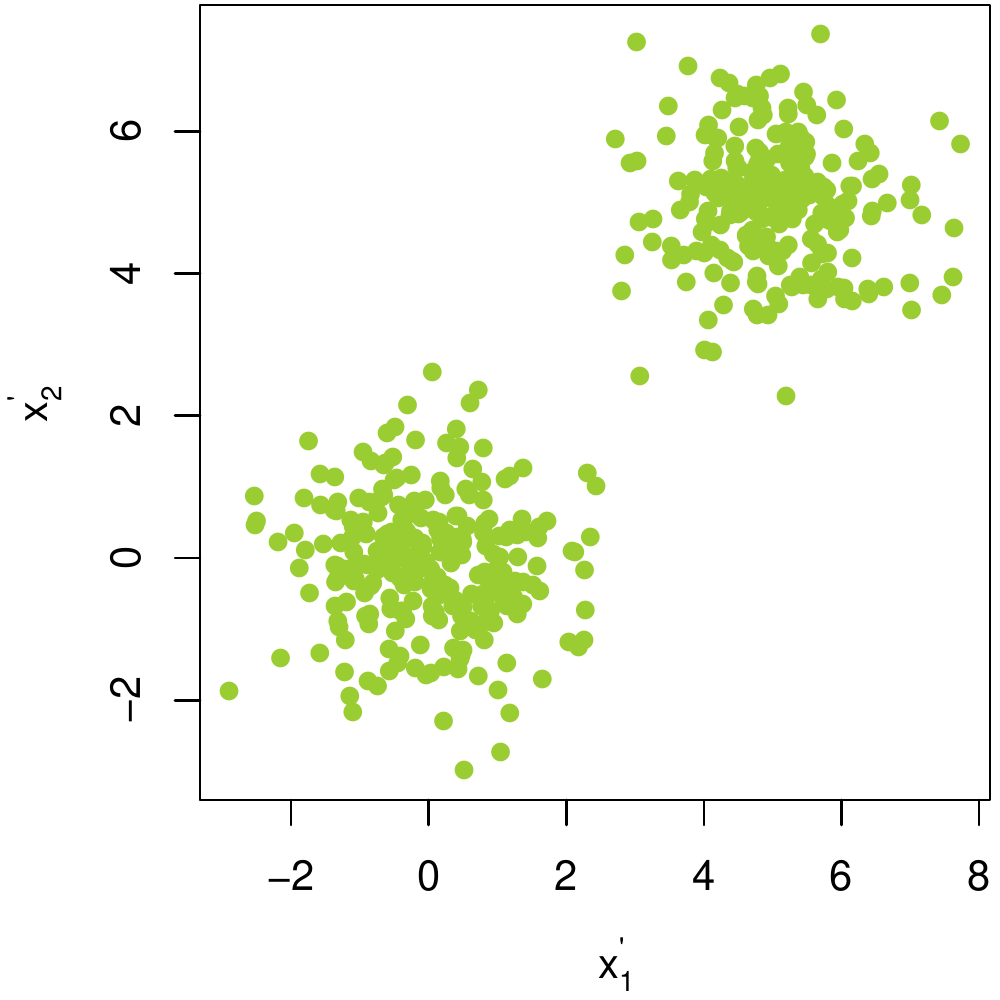}
 }

    \caption[EDA.]{FGMO example.
      \subref{figFgmA} The first iteration of the EM algorithm. The original data slightly overlapped with two mixture components.
      The red dots represent the means of each mixture and the contour curves their densities,
    \subref{figFgmB} the last iteration of the EM and
    \subref{figFgmC} the sampled values.}%
    \label{figFgm}%
\end{figure*}

Figure \ref{figFgm} shows an example of how the FGMO works. At the first iteration, the parameters are set to the initial condition.
Then, according to the EM iterations, the parameters tend to converge. Figure~\ref{figFgmB} shows the parameters after the convergence.
Finally, new values are sampled (Figure~\ref{figFgmC}) from its fitted mixtures model.

\section{Results}
\label{results}

The results show that the three proposed EDAs (UNIO, KDEO and FGMO) performed properly for PSP. For each proposed method, we evaluated
three different aspects: (1) the computational cost, measuring the overall time of a prediction; (2) the van der Waals energy value of
the best individual at the last generation and (3) the RMSD, which represents how similar a solution is to the native protein.

We have to be careful when evaluating (3) since we are only using van der Waals energy in our fitness function. In this case, we are
bypassing other potentials that are less relevant to stabilize the molecule. Adding other potentials in pure \emph{ab initio} PSP would
require a much more complex algorithm, i.e. the Multi-Objective method proposed by \cite{Brasil2013}. This paper shows that we can find
adequate pure \emph{ab initio} protein configurations using only van der Waals and the proposed EDAs. We compared the KDEO against other
optimization methods from the literature, as Random Walk (RW) \cite{Pearson1905}, Monte Carlo (MC) \cite{Metropolis1949}, Genetic
Algorithm (GA) \cite{Goldberg2002} and Differential Evolution (DE) \cite{Storn1997}. Appendix A shows a pairwise comparison using
the Wilcoxon test \cite{Hollander1973} for all evaluated methods considering the three aspects: van der Waals energy, RMSD and running time.

\subsection{Experimental setup}

The experiments were run in the 20 nodes cluster of the Laboratorio de Computacao Reconfiguravel at the ICMC-USP. Each node has an
Intel Core i7 2.67~GHz processor with 4 physical processors, 8 considering Hyper-Threading technology. It also has 32 GB of RAM
and the Operational System is Debian 4.6.3-14 64 bits. Moreover, each node has two network adapters. One is used for the file
system (NFS) and other to communication operations in MPI \cite{Quinn2004}.

\begin{figure}%

    \centering
      \subfloat[][]{%
        \label{natives}%
        \includegraphics[width=.2\textwidth]{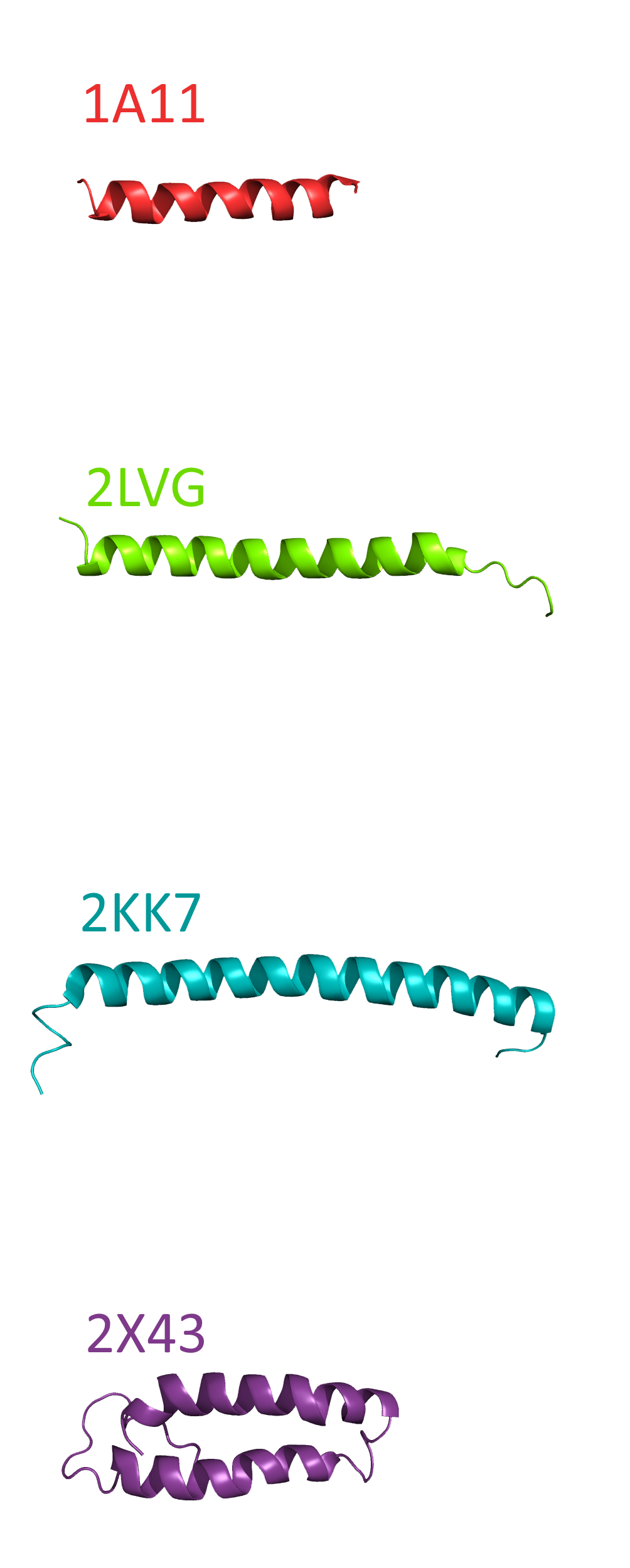}
 }
    \subfloat[][]{%
        \label{ppDensity}%
        \includegraphics[width=.6\textwidth]{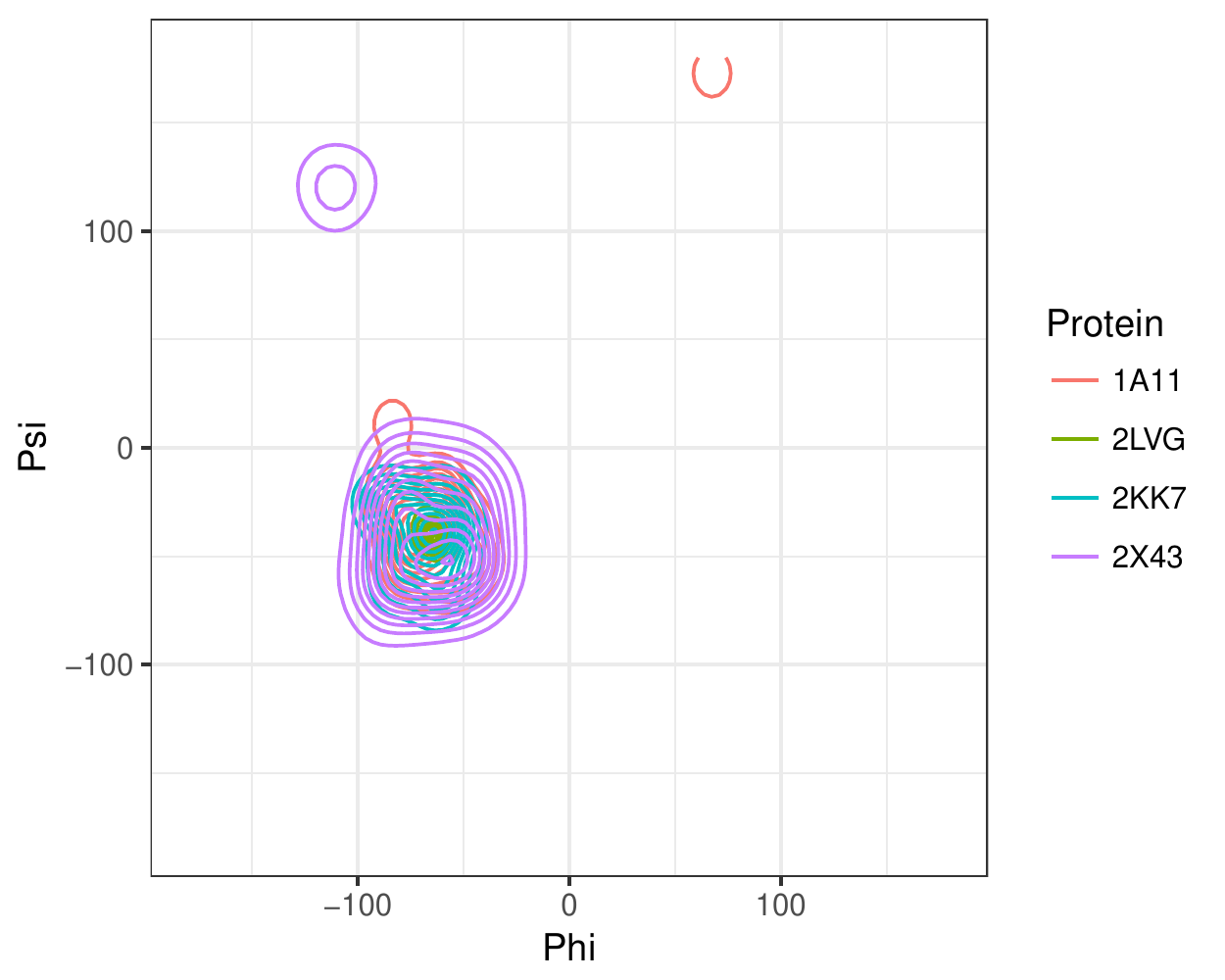}
 }

    \caption[EDA.]{Native proteins:
    \subref{natives} three dimensional structures and,
    \subref{ppDensity} the distribution of the $[\phi;\psi]$ dihedral angles of the corresponding proteins.}%
    \label{nativeProteins}%
\end{figure}

We have selected four small proteins from the Protein Data Bank (PDB) \cite{PDB2009}, all containing $\alpha$-helices. This may be
favorable for the optimization model used that is based on van der Waals only. Table~\ref{tabProts} shows the PDB ID, the number of
residues and the problem length ($m$). Figure~\ref{nativeProteins} shows the shape and the $[\phi, \psi]$ dihedral angles distributions
of each native structure.

\begin{table}[htbp]
  \centering
  \caption{Protein sizes in experiments.}
    \begin{tabular}{rrr}
    Protein & Residues & Problem length \\
    1A11  & 25    & 95 \\
    2LVG  & 40    & 169 \\
    2KK7  & 52    & 229 \\
    2X43  & 67    & 268 \\
    \end{tabular}%
  \label{tabProts}%
\end{table}%

The convergence is reached when either each of the tested algorithms reaches one million evaluations or the standard deviation
of the population fitness falls below $0.0001$.

\subsubsection{Performance issues}

ProtPred is entirely written in C language and it uses efficient libraries as GSL and CBLAS \cite{GSL} to deal with most  algebraic and
statistical operations. Some functions were taken from statistical language R \cite{RCT2014} and translated into C. The van der Waals
energy uses an efficient implementation based on cell-lists, as we proposed in an earlier work \cite{Bonetti2013}. Furthermore, the
tested approaches have several different parameters to set, so we used MPI in order to distribute each algorithm configuration
throughout our cluster. Despite this, we needed about 12,000 hours of CPU time to get all the results shown in this paper.

\subsection{Proposed Estimation of Distribution Algorithms}
\label{prop}

Before we run the sequence of experiments, we found a set of parameters of the EDA according to the probabilistic model used (Table~\ref{set1}).
For each combination of parameters, the experiments were repeated 30 times with different seeds for each protein.

\begin{table}[htbp]
  \centering
  \caption{Parameters used in experiments with the probabilistic models.}
    \begin{tabular}{ccccc}
    \toprule
          & Pop.  & Selected  & Tournament & Mixtures \\
          & size  & size       & size & \\
    \midrule
    UNIO  & 2000  & 2000  & 2     & -  \\
    KDEO  & 500   & 250   & 2     & -  \\
    FGMO  & 200   & 100   & 2     & 10 \\
    \bottomrule
    \end{tabular}%
  \label{set1}%
\end{table}%

\begin{figure*}%

    \subfloat[][]{%
        \label{c1a11a}%
         \includegraphics[width=.35\textwidth]{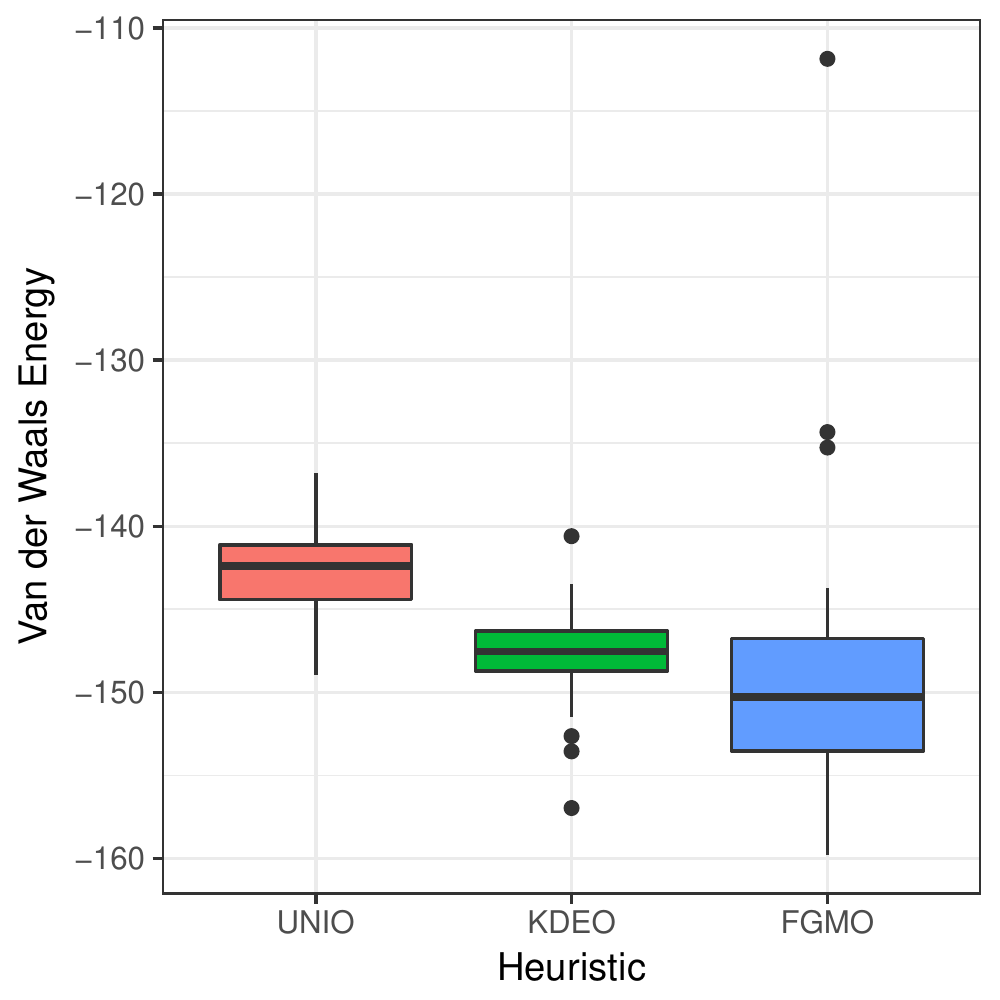}
 }
        \hspace{8pt}%
    \subfloat[][]{%
        \label{c1a11b}%
        \includegraphics[width=.35\textwidth]{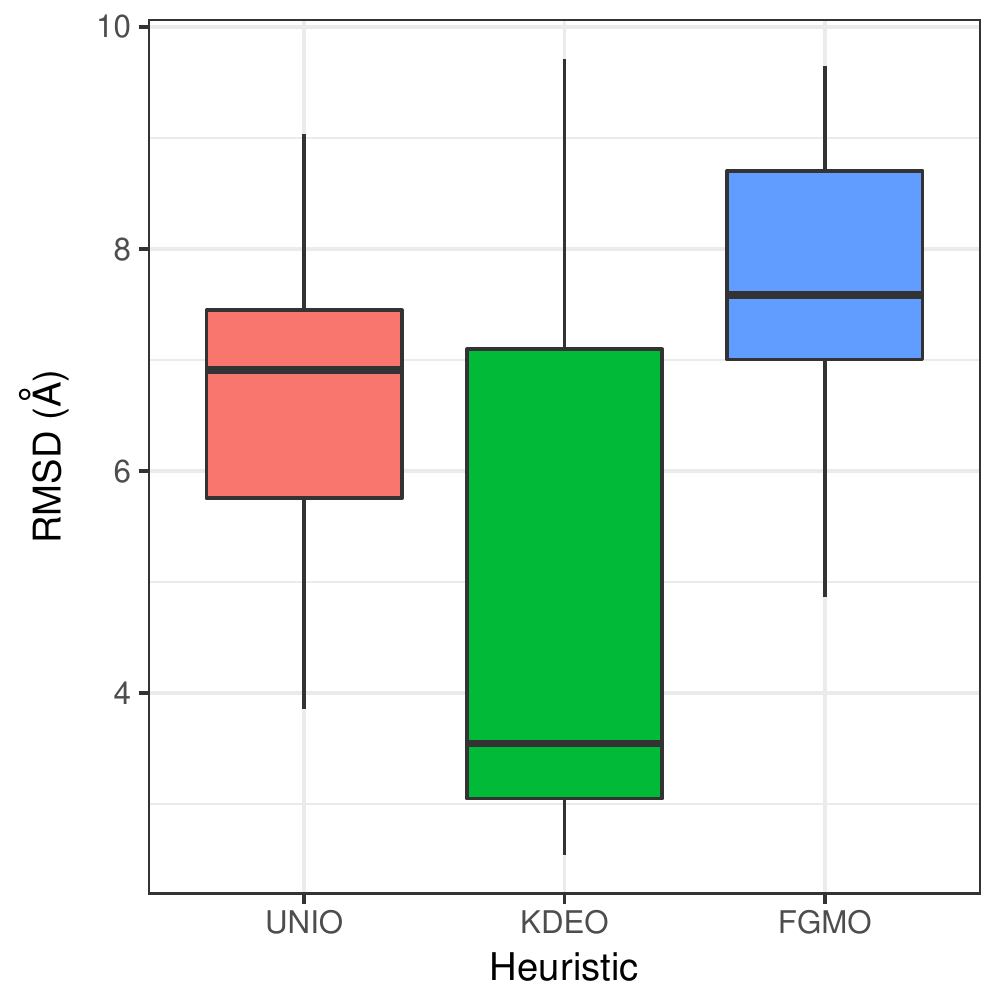}
 }\\
    \subfloat[][]{%
        \label{c1a11c}%
        \includegraphics[width=.35\textwidth]{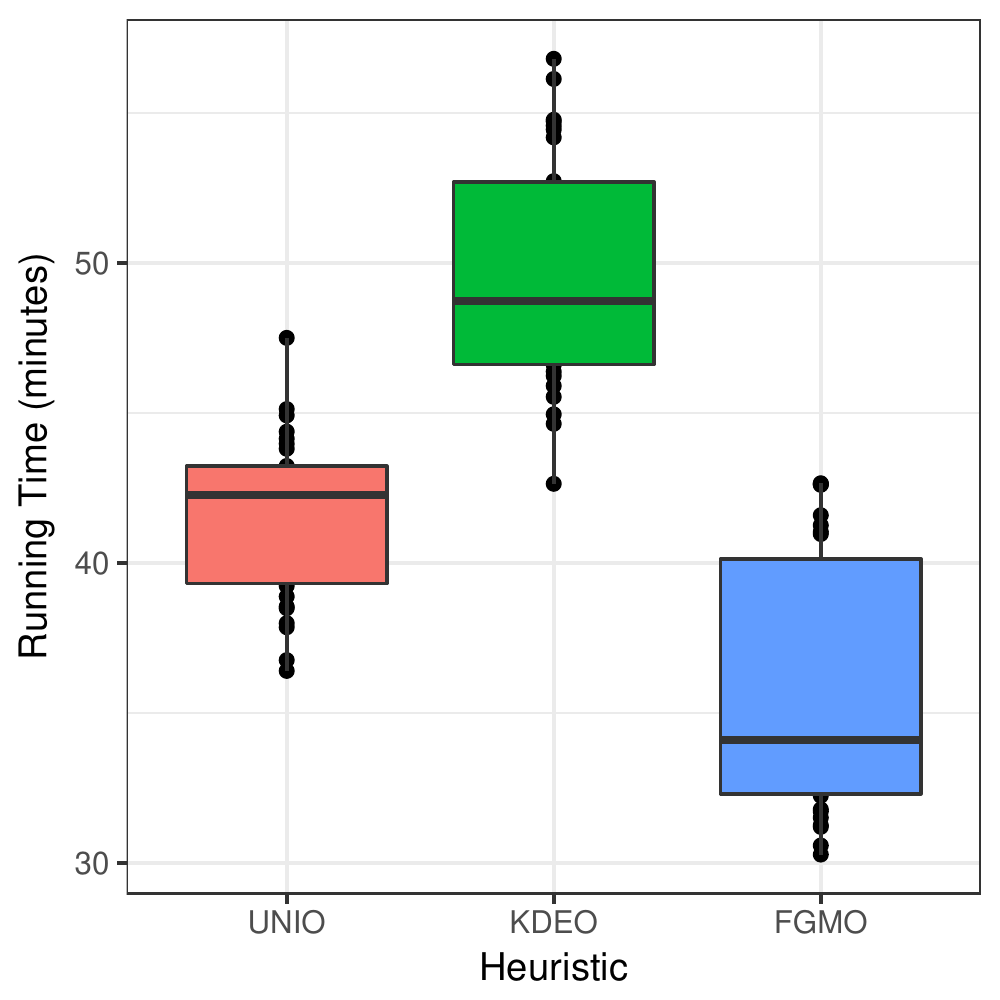}
 }
     \hspace{8pt}%
    \subfloat[][]{%
        \label{c1a11d}%
        \includegraphics[width=.44\textwidth]{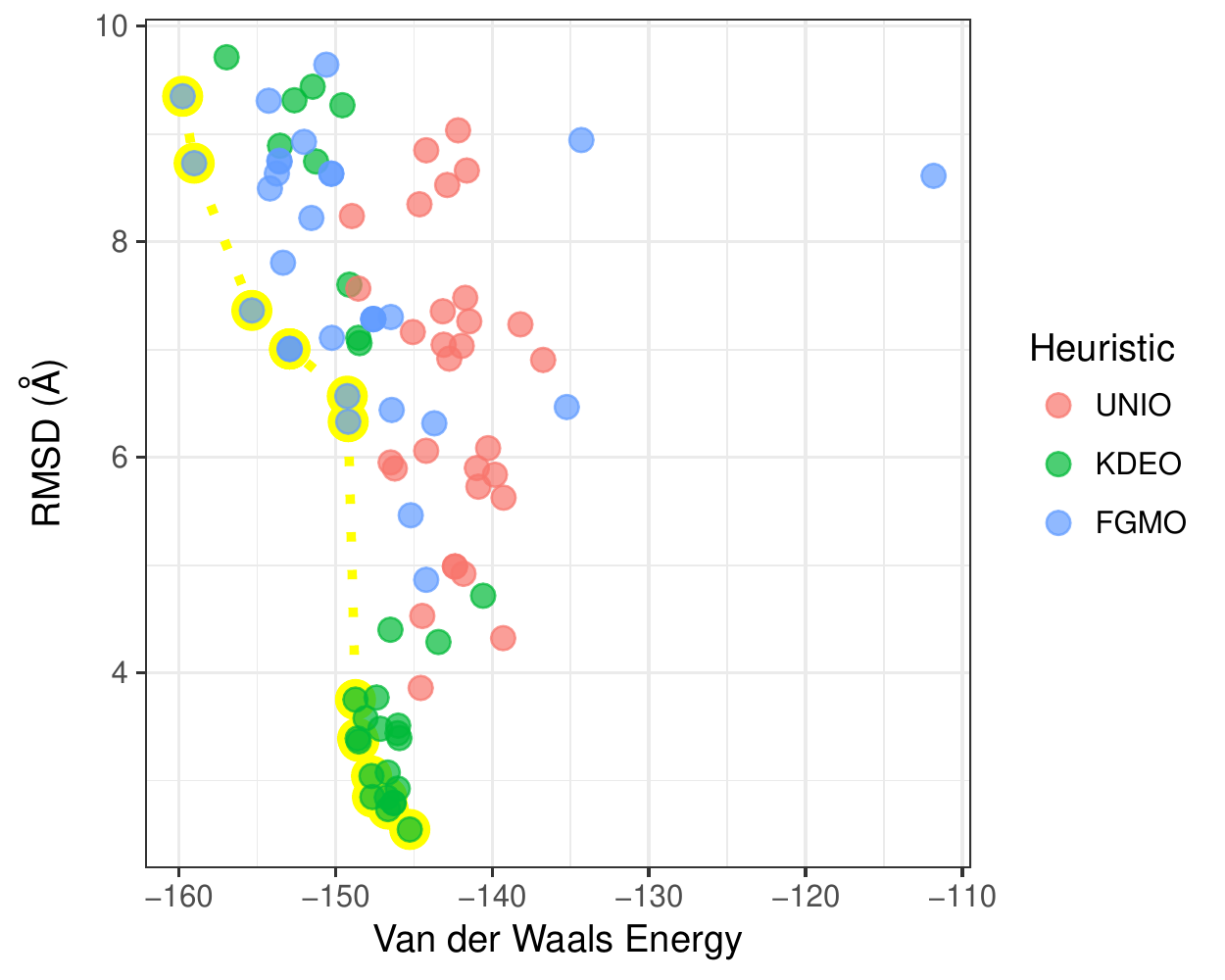}
 }

    \caption[EDA.]{EDA for protein 1A11 with the three proposed methods:
    \subref{c1a11a} energy plot;
    \subref{c1a11b} RMSD plot;
    \subref{c1a11c} time needed to run; and,
    \subref{c1a11d} the scatter plot between energy and RMSD and the Pareto Front highlighted in yellow.}%
    \label{c1a11}%
\end{figure*}

Figure~\ref{c1a11} shows the results obtained for the smaller protein 1A11. Considering the energy aspect only, the FGMO was best,
although it has some high energetic outliers. The KDEO performed best when considering the RMSD. The running time results were
partially surprising since a sophisticated probabilistic model as FGMO was faster than UNIO, in most  cases. That happens because
the selected size used by UNIO needed to be 20 times larger than FGMO.

%\overfullrule=1mm

% Descomentar
In the PSP problem, a trade-off between energy and RMSD is sometimes difficult to obtain since the predicted energy can exceed the
native energy value, when using only van der Waals energy. A scatterplot of van der Waals energy per RMSD (Figure \ref{c1a11d} for
protein 1A11) enables to evaluate such trade-off. We also highlighted in yellow the set of points that dominate other ones, in terms
of multi-criteria optimization \cite{Srinivas1994}. These solutions in yellow approximate a Pa\-re\-to Front. The points from UNIO
do not show up in the Front, meaning that neither the RMSD nor the energy aspects were better than FGMO or KDEO.

% 1A11

\begin{figure*}%

    \subfloat[][]{%
        \label{c2lvga}%
       \includegraphics[width=.4\textwidth]{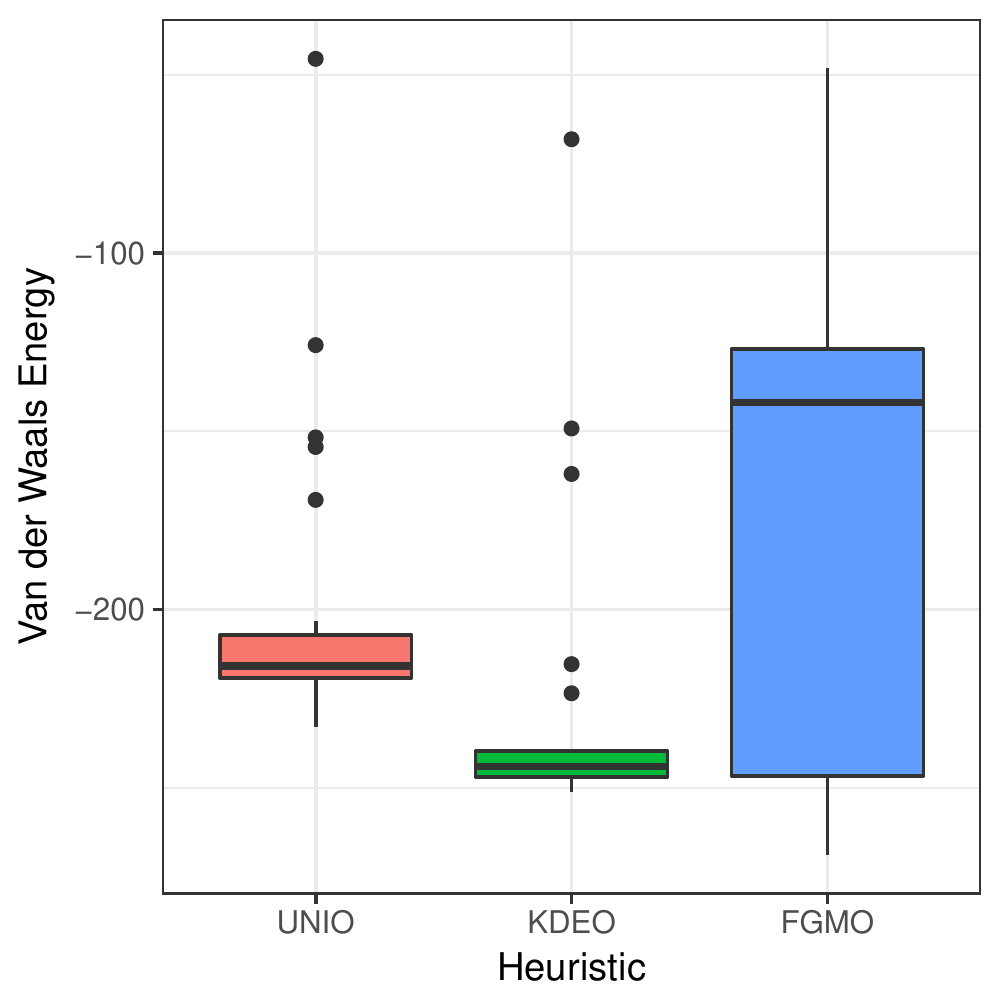}
}
        \hspace{8pt}%
    \subfloat[][]{%
        \label{c2lvgb}%
       \includegraphics[width=.4\textwidth]{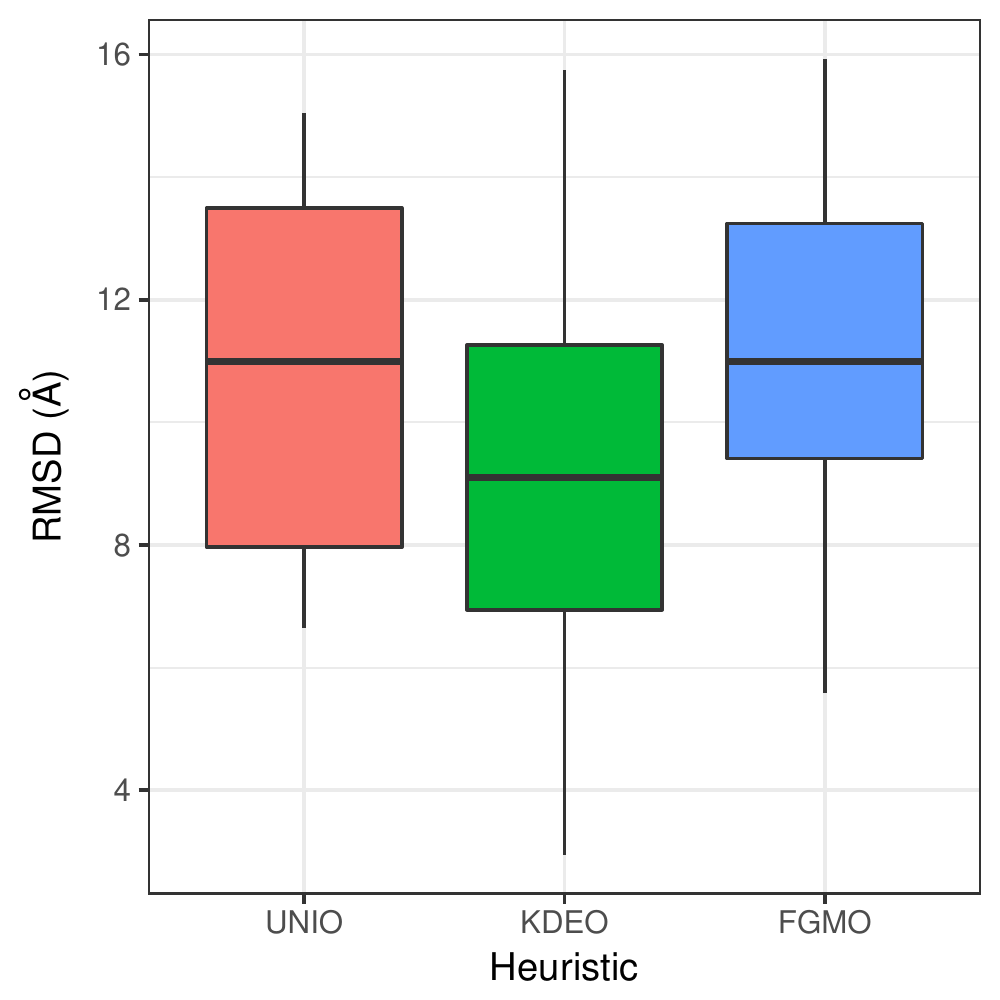}
 }\\
    \subfloat[][]{%
        \label{c2lvgc}%
        \includegraphics[width=.4\textwidth]{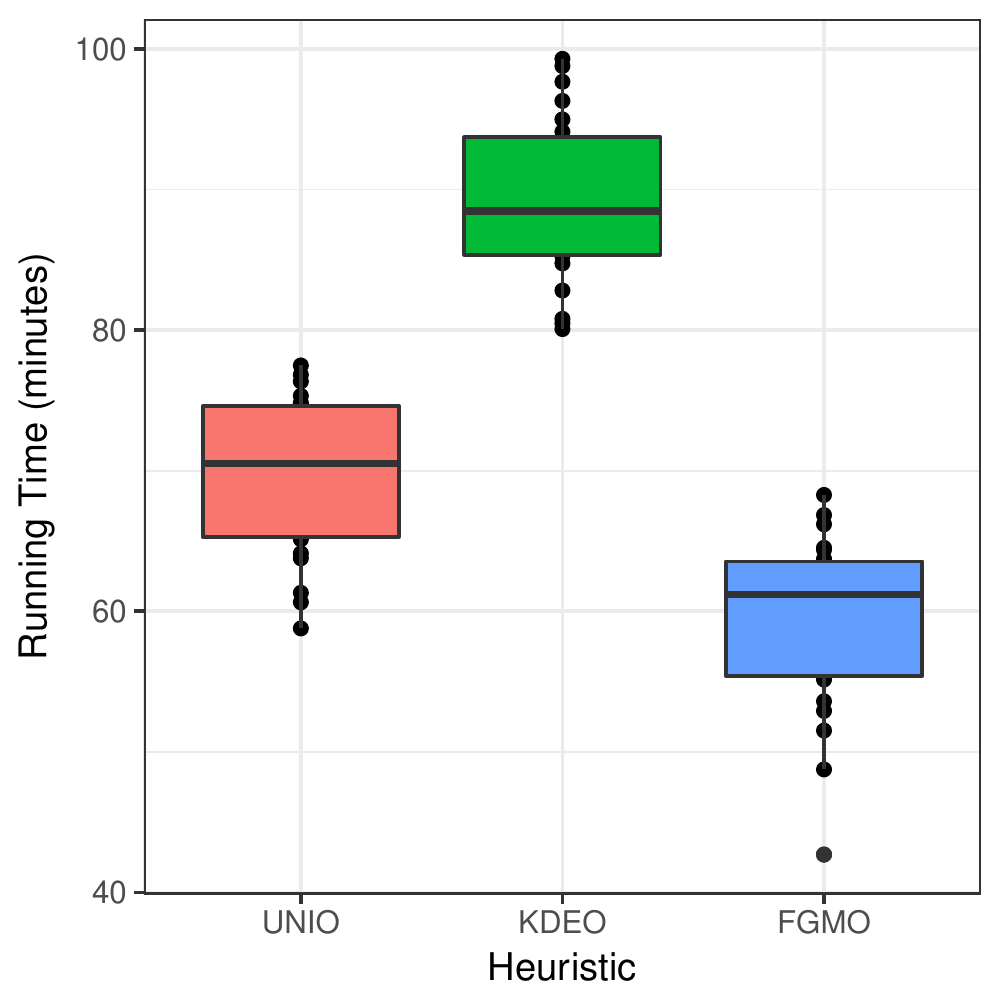}
 }
     \hspace{8pt}%
    \subfloat[][]{%
        \label{c2lvgd}%
        \includegraphics[width=.5\textwidth]{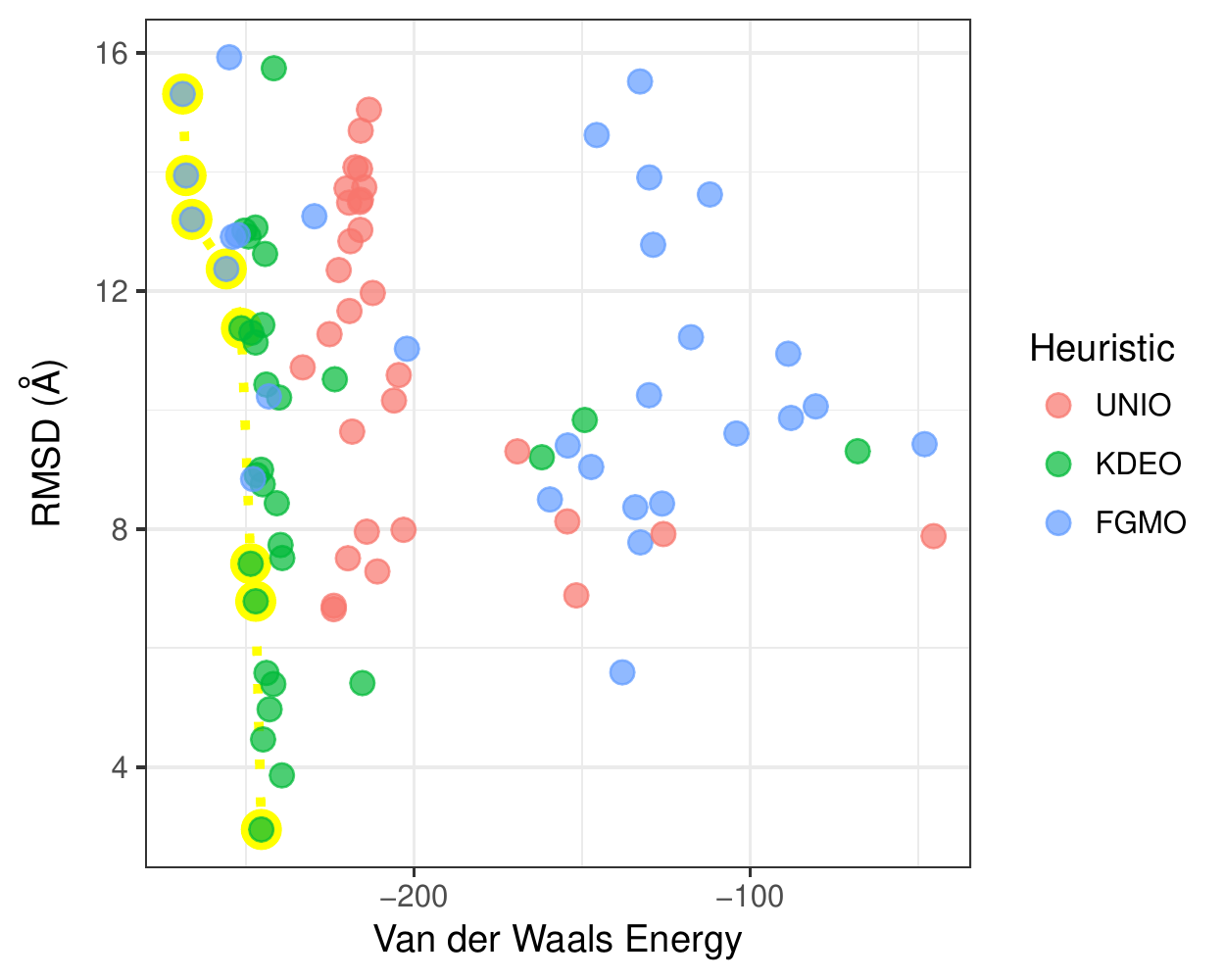}
 }

    \caption[EDA.]{EDA for protein 2LVG with the three proposed methods:
    \subref{c2lvga} density plot of the energy;
    \subref{c2lvgb} RMSD plot;
    \subref{c2lvgc} time needed to run; and,
    \subref{c2lvgd} the scatter plot between energy and RMSD and the Front highlighted in yellow.}%
    \label{c2lvg}%
\end{figure*}

The results for the protein 2LVG with 40 residues are shown in Figure~\ref{c2lvg}. Although FGMO obtained a large variance in the energy,
it also got the smallest values found, while UNIO and KDEO had concentrated points. KDEO was the only method capable of finding RMSD
values below 5.0 and reached an average better than UNIO and FGMO as well. The running time followed the same pattern as the 1A11
protein. KDEO was the slower among the three and FGMO was the faster.

In the scatterplot of Figure~\ref{c2lvgd}, most of the left upper points in the Front belong to FGMO with smallest energy, and the
left lower points belongs to KDEO. This means FGMO was able to minimize better than KDEO, but KDEO found solutions closest to the
configuration found in nature.

% 2LVG

\begin{figure*}%

    \subfloat[][]{%
        \label{c2kk7a}%
         \includegraphics[width=.4\textwidth]{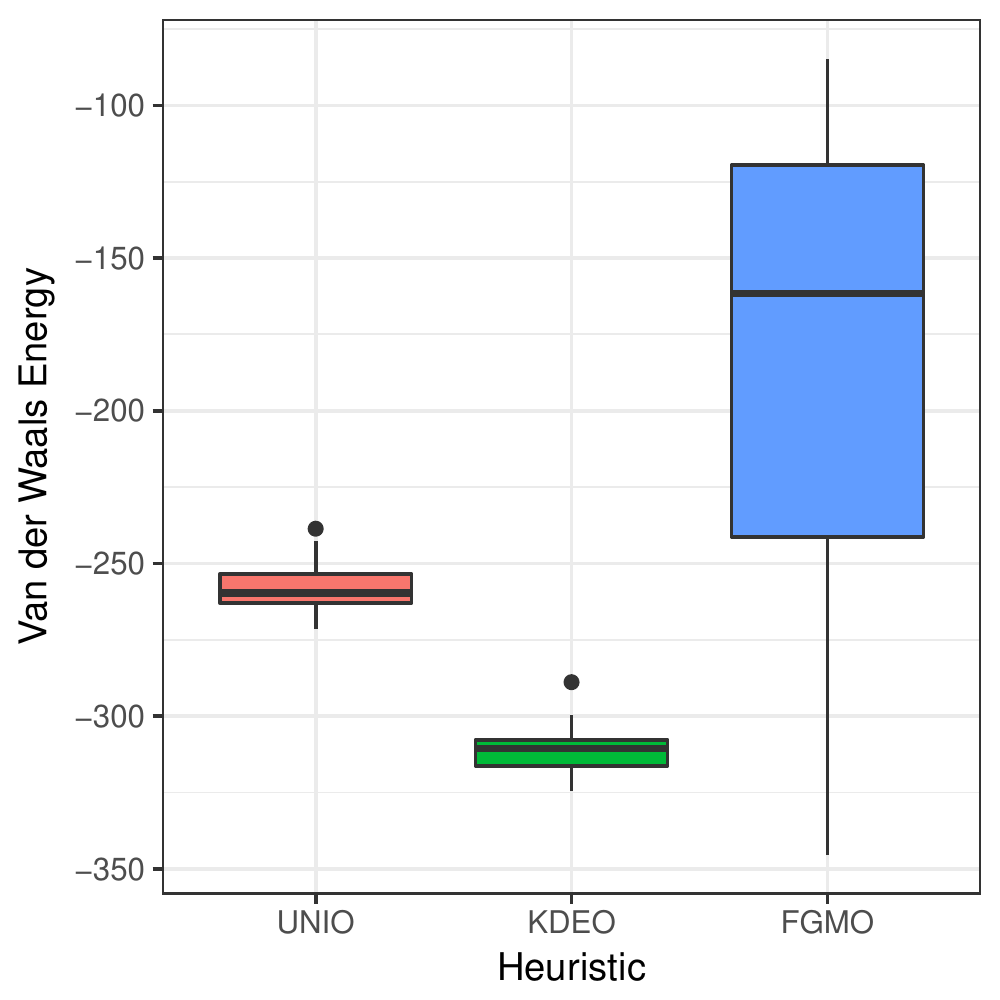}
 }
        \hspace{8pt}%
    \subfloat[][]{%
        \label{c2kk7b}%
        \includegraphics[width=.4\textwidth]{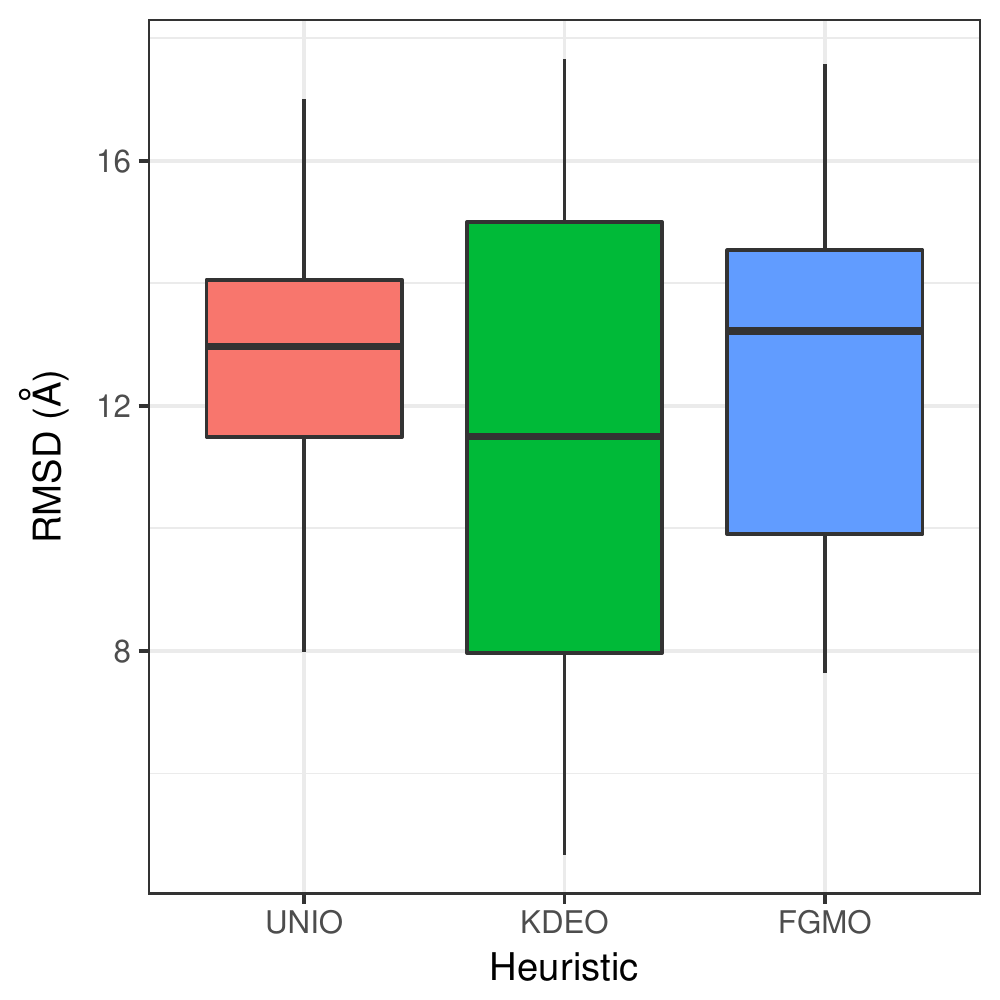}
 }\\
    \subfloat[][]{%
        \label{c2kk7c}%
        \includegraphics[width=.4\textwidth]{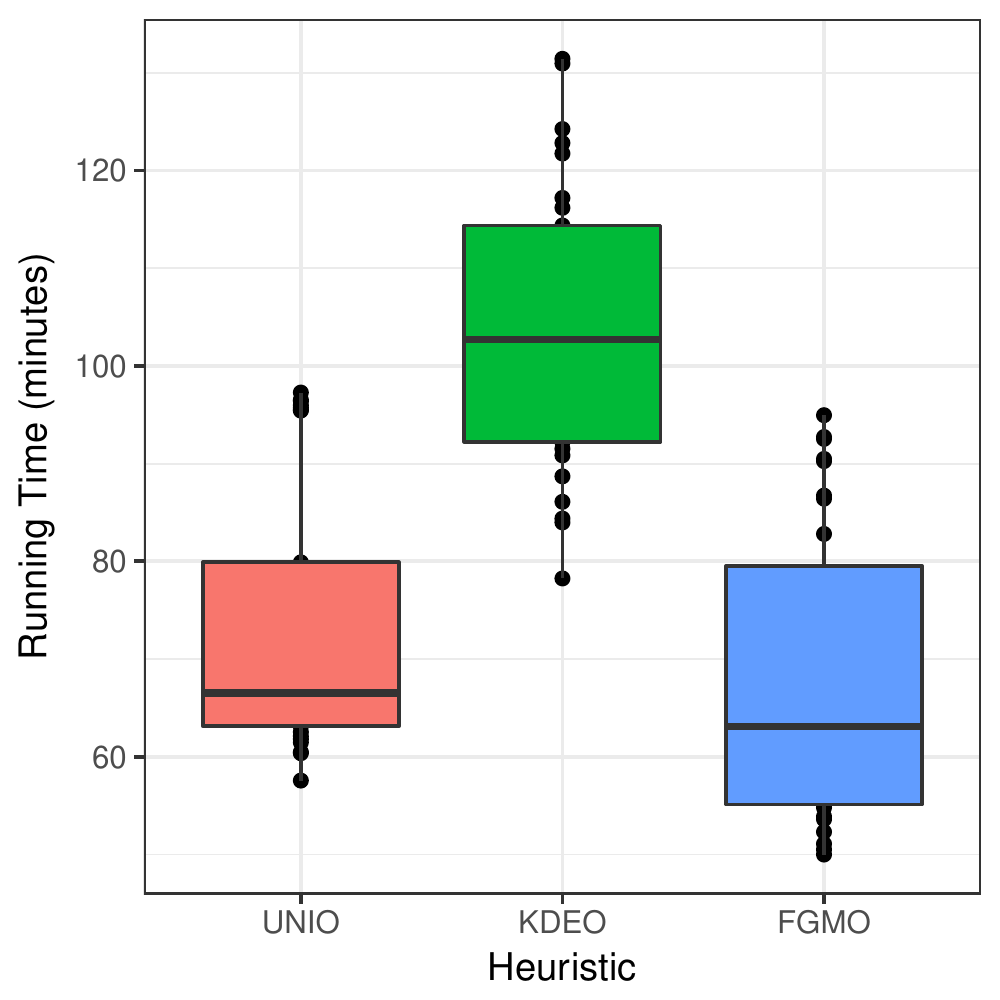}
 }
     \hspace{8pt}%
    \subfloat[][]{%
        \label{c2kk7d}%
        \includegraphics[width=.5\textwidth]{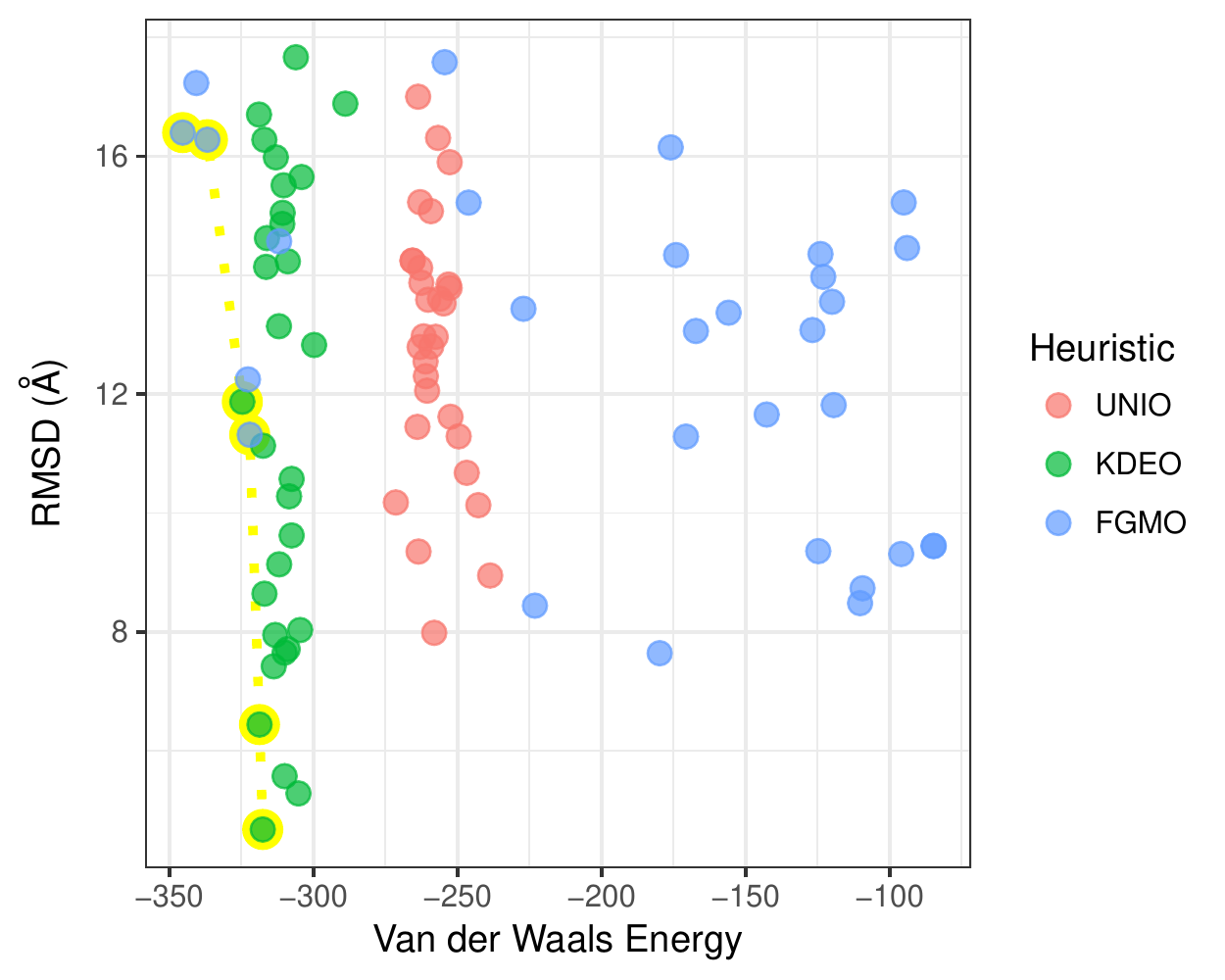}
 }

    \caption[EDA.]{EDA for protein 2KK7 with the three proposed methods:
    \subref{c2kk7a} density plot of the energy;
    \subref{c2kk7b} RMSD plot;
    \subref{c2kk7c} time needed to run; and,
    \subref{c2kk7d} the scatter plot between energy and RMSD and the Pareto Front highlighted in yellow.}%
    \label{c2kk7}%
\end{figure*}

For protein 2KK7 with 52 residues (Figure~\ref{c2kk7}), FGMO also had widely-spaced energy values. Considering only the
average measure, FGMO would be the worst, but it was able to find the best energy value. It means that FGMO can reach
better solutions than others, but for some reason, it is getting stuck at some local optimum, worsening the average
value. Considering the RMSD aspect, KDEO was the only one able to find solutions with RMSD below $7.0$. The running
times required by UNIO and FGMO become closer due to fact that this protein is larger than 1A11 and 2LVG are. More
points belonging to KDEO appear in the van der Waals energy per RMSD scatterplot. The top left points in the Front
belong to FGMO. They had the best energy but an inferior RMSD.

% 2KK7

\begin{figure*}%

    \subfloat[][]{%
        \label{c2x43a}%
         \includegraphics[width=.4\textwidth]{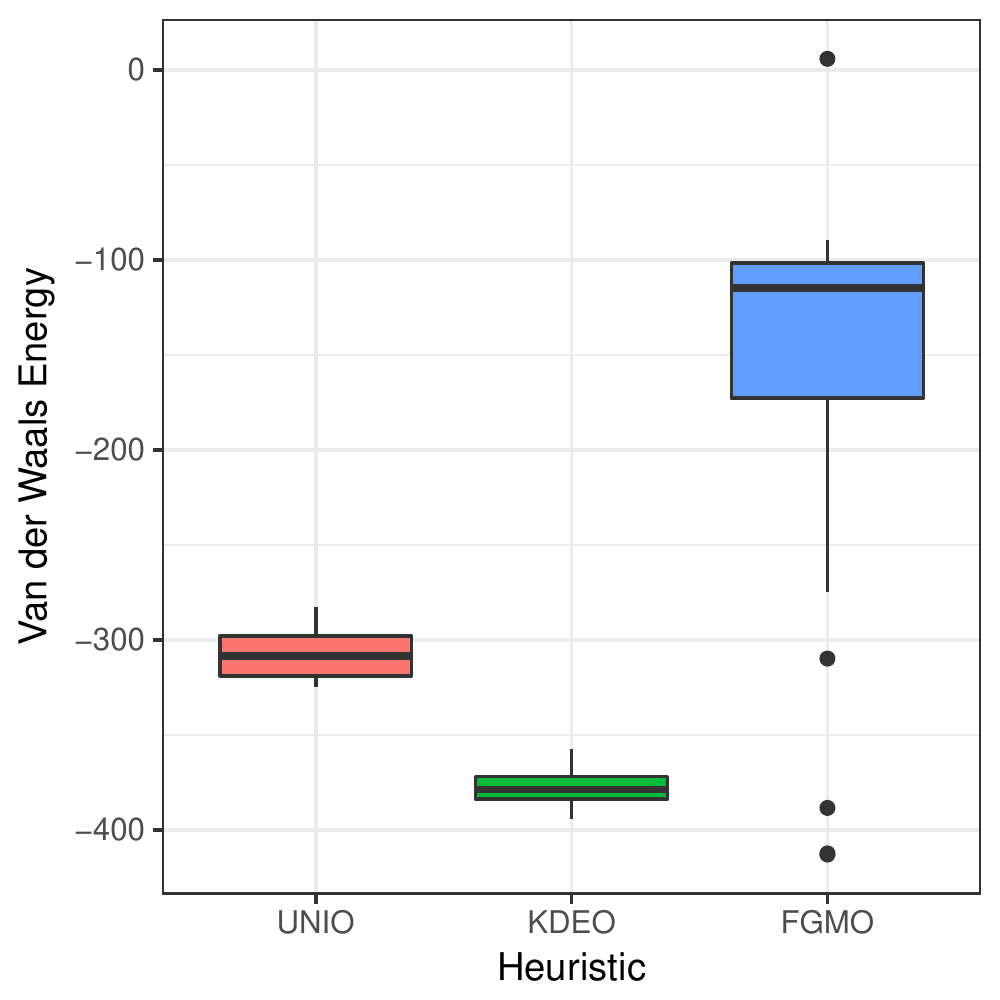}
 }
        \hspace{8pt}%
    \subfloat[][]{%
        \label{c2x43b}%
        \includegraphics[width=.4\textwidth]{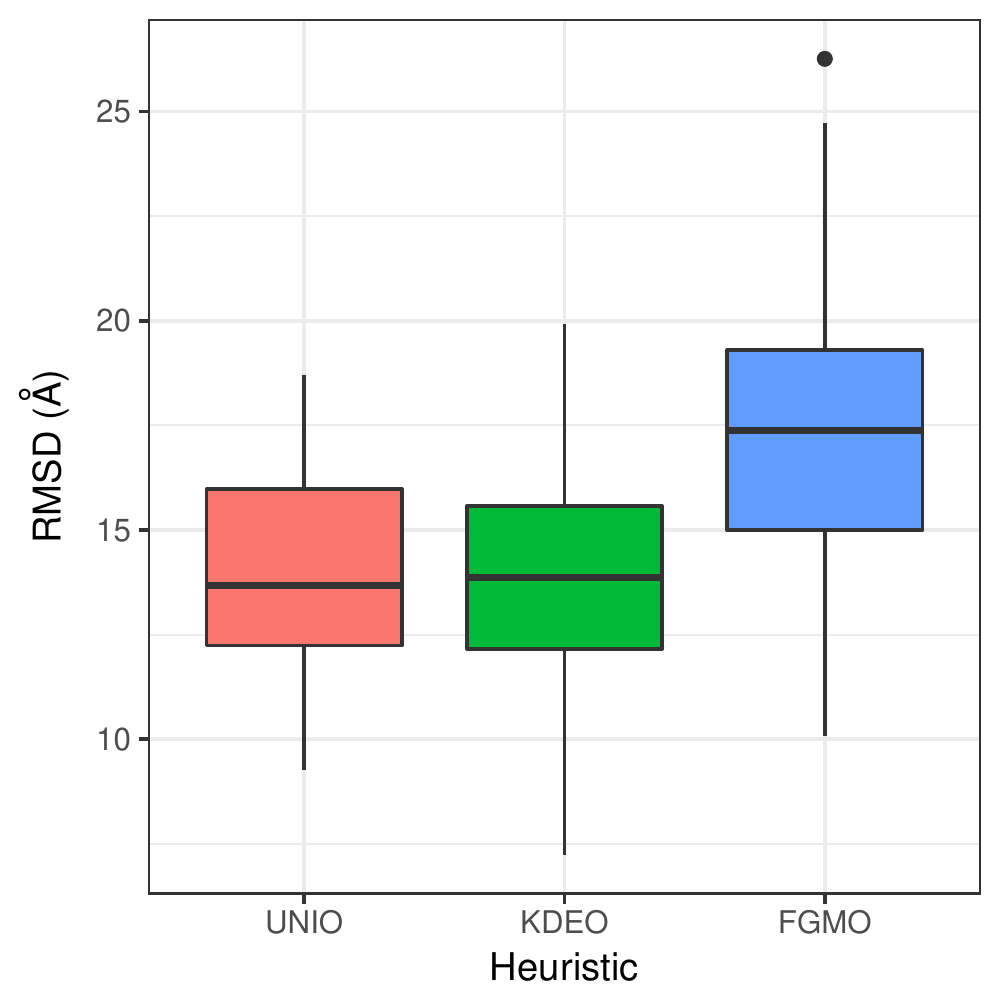}
 }\\
    \subfloat[][]{%
        \label{c2x43c}%
        \includegraphics[width=.4\textwidth]{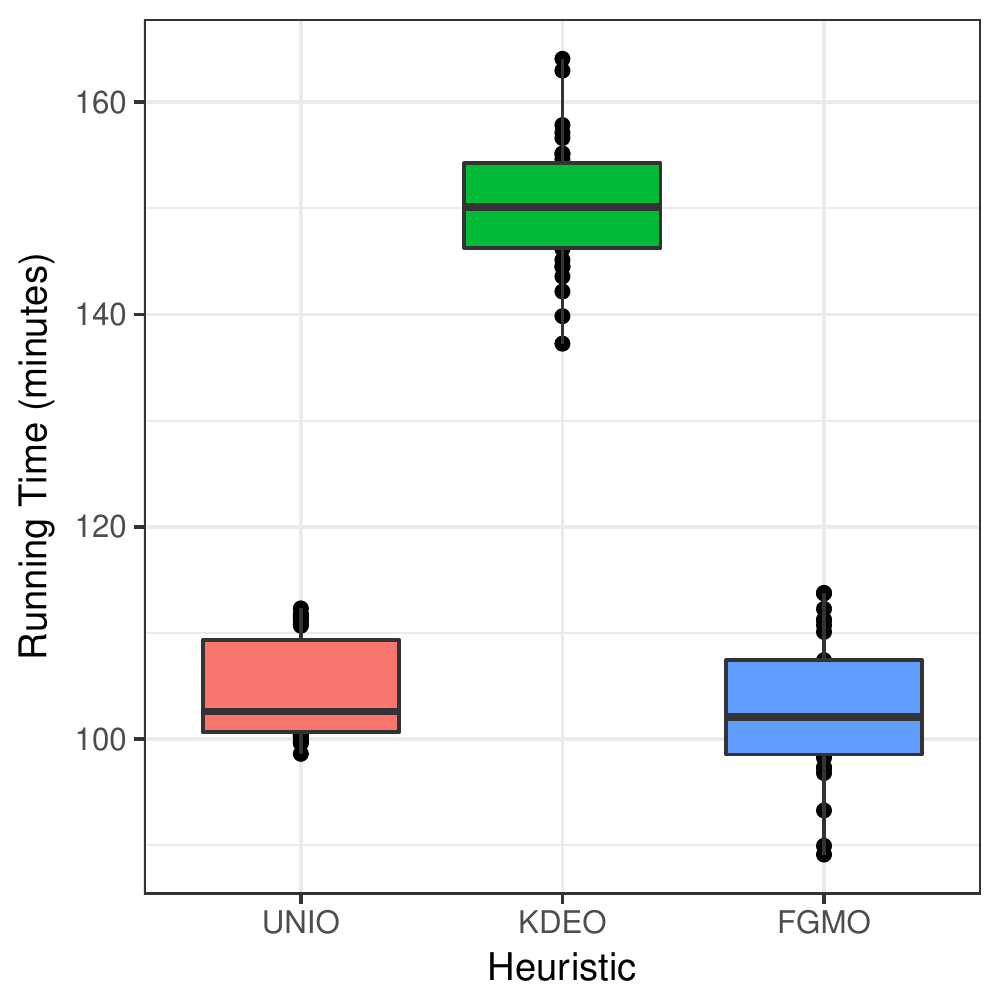}
 }
     \hspace{8pt}%
    \subfloat[][]{%
        \label{c2x43d}%
        \includegraphics[width=.5\textwidth]{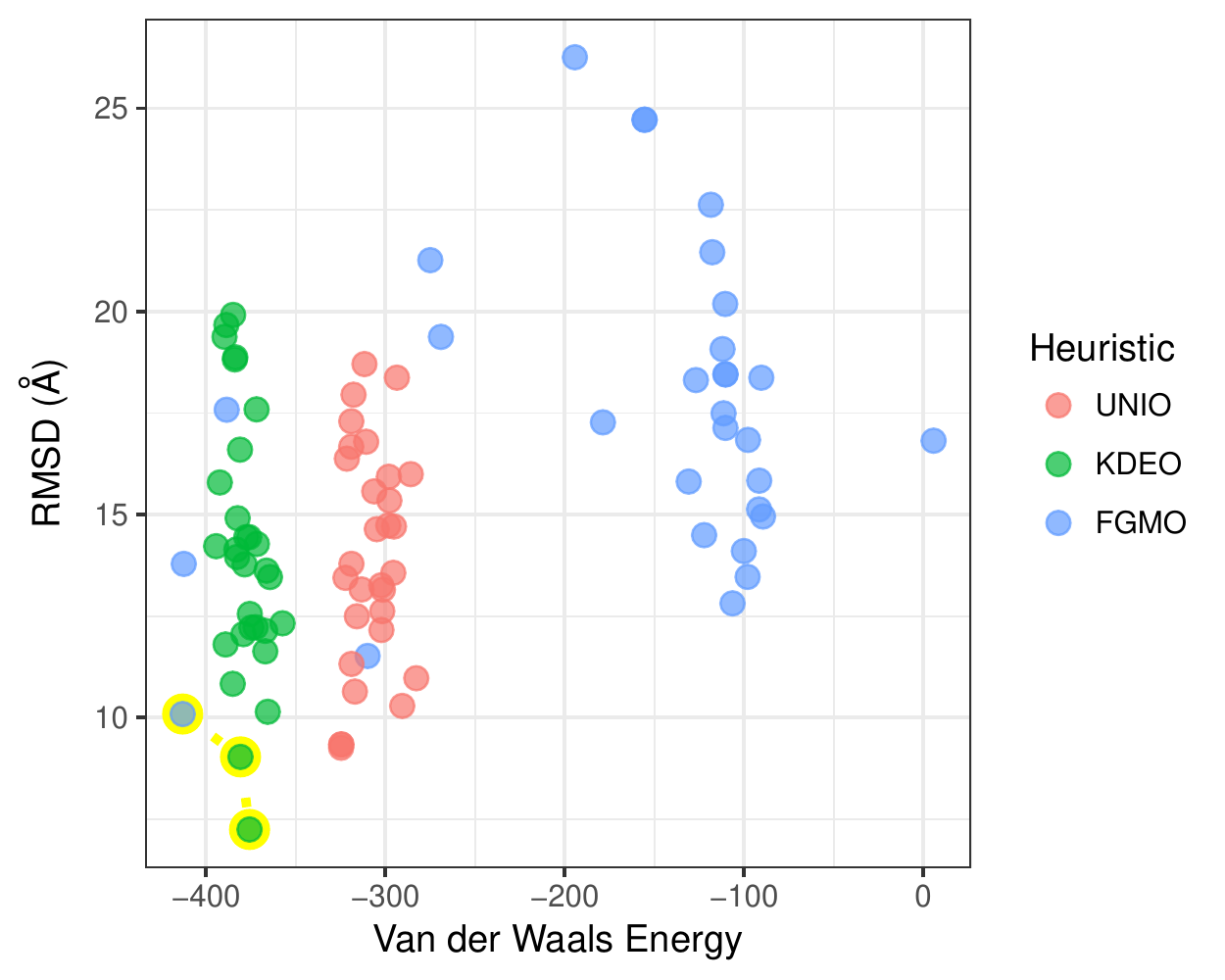}
 }

    \caption[EDA.]{EDA for protein 2X43 with the three proposed methods:
    \subref{c2x43a} density plot of the energy;
    \subref{c2x43b} RMSD plot;
    \subref{c2x43c} time needed to run; and,
    \subref{c2x43d} the scatter plot between energy and RMSD and the Pareto Front highlighted in yellow.}%
    \label{c2x43}%
\end{figure*}

Finally, the results for the largest protein used in the experiments, 2X43 with 67 residues follow the same trend as the
previous proteins (Figure~\ref{c2x43}). Considering the energy aspect only, the UNIO and KDEO were, in average, better than
FGMO is. However, Figure~\ref{c2x43a} shows that two outliers of FGMO are below $-410$, i.e. less than all other points from
UNIO and KDEO. We believe that happened due to the sensitiveness of FGMO when setting the same number of mixture components
throughout the evolutionary process, which somehow was benefited by the entirely evolutionary process. Apart from the KDEO
running time, UNIO and FGMO had similar results. It seems that the running time of FGMO would be higher than UNIO for large
proteins.

The scatterplot from Figure~\ref{c2x43d} shows a more defined agglomeration of points. KDEO produced two points in the Front.
On the other hand, FGMO returned one point in the Front, although most of the points are around $-300$ of energy with RMSD between $15$ and $25$.

% 2X43

The Table~\ref{tabFinalTable} shows a summary of all experiments for all the four proteins evaluated. We can see that FGMO was the
algorithm that got the best energy values; the KDEO the best protein structures; and UNIO the best running times. The KDEO also got the
worse running time among all evaluated algorithms. However, it is interesting to see that a relatively complex algorithm as the proposed
FGMO was faster than RW for most of proteins evaluated.

\input{appendix/stat-tabFinalTable.tex}

\subsection{Comparison with the Native energy}

We compared all non-bonded energies (van der Waals, electrostatic, solvation and hydrogen bonding) of each of the
best predicted protein configuration against the native structure of each used protein. In order to compute the
energy of native proteins, we converted the XYZ Cartesian coordinates from the PDB file into dihedrals and then
used these values as input of one fitness function call of ProtPred.

\begin{figure*}%
    \centering
    \includegraphics[width=.9\textwidth]{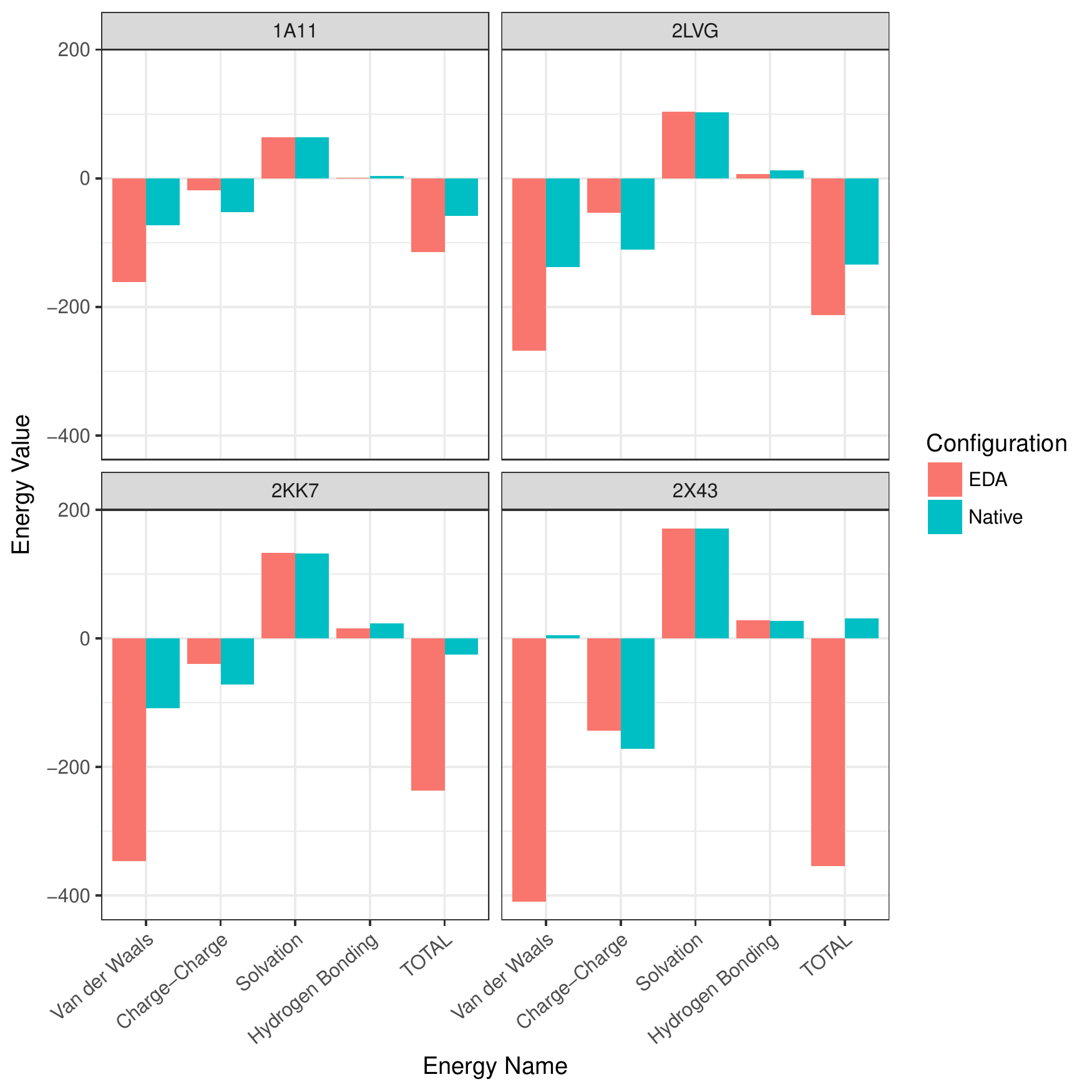}
    \caption[EDA.]{Comparison of the individual energies between the proposed EDA and the native protein configuration for proteins 1A11, 2LVG, 2KK7 and 2X43.
    \label{energiesNativeEDA}}
\end{figure*}

Figure~\ref{energiesNativeEDA} shows a comparison between the configuration with the best energy value found by the
proposed EDA and the native configuration. For all the four proteins evaluated, the EDA managed to minimize the van
der Waals energy more than native energy. This was expected because only van der Waals energy was minimized. However,
in native proteins, the electrostatic energy (charge-charge) was better than the structure found by the proposed EDA.
Thus, if we attempt to decrease the electrostatic energy of the protein configuration obtained by our EDA, the van der
Waals energy would probably increase. The solvation and hydrogen bonding energies of the EDA were also higher than the
native were, apart from the hydrogen bonding of protein 2X43. Therefore, a multi-objective algorithm would be a more
appropriate strategy in order to deal correctly with all these energies as proposed by \cite{Brasil2013}.

\pagebreak

\subsection{Comparison with other Evolutionary Algorithms}

We showed that FGMO can produce more diversified solutions, yielding promising energy values. However, KDEO was able to
find a better compromise between energy and RMSD. For this reason, we selected the KDEO to be compared against other
heuristics from the literature. The parameters used with the other heuristics are defined in Table~\ref{tabHeur}.

\begin{table}[htbp]
  \centering
  \caption{Parameters used with RW, MC, GA and DE.}
    \begin{tabular}{rrrrr}
    \toprule
          & RW    & MC    & GA    & DE \\
    \midrule
    Pop. Size & 100   & 1000  & 2000  & 1000 \\
    Sel. Pressure &       &       &       & 1.2 \\
    Tournament &       &       &       &  \\
    Step  &       & 2     &       &  \\
    Cross. Rate &       &       & 0.3   & 0.2 \\
    Mutation Rate &       &       & 0.5   &  \\
    Mutation Factor &       &       & 0.5   &  \\
    Differential weight &       &       &       & 0.2 \\
    \bottomrule
    \end{tabular}%
  \label{tabHeur}%
\end{table}%

\begin{figure*}%

    \centering
    \subfloat[][]{%
        \label{ch1a11}%
         \includegraphics[width=.45\textwidth]{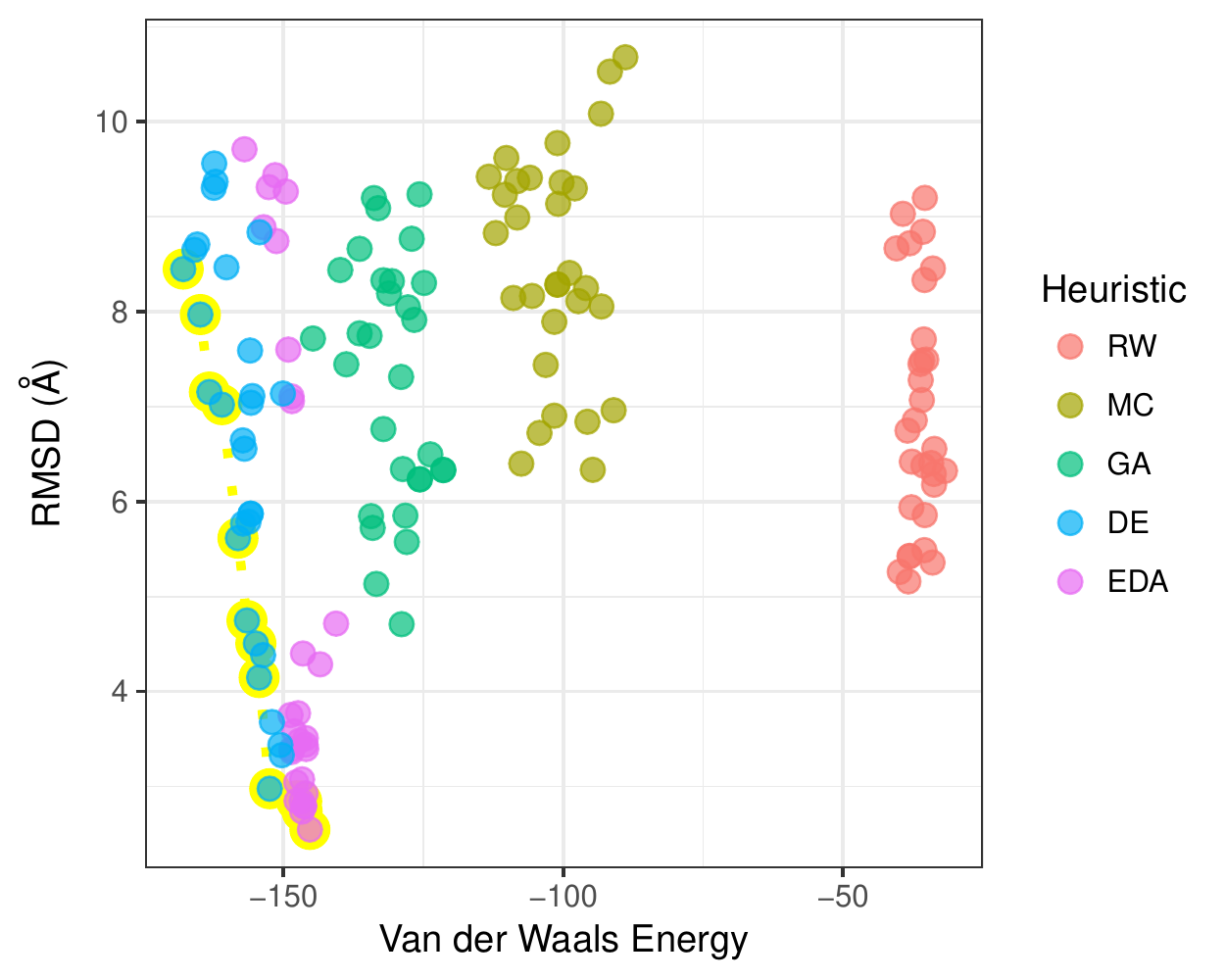}
 }
    \subfloat[][]{%
        \label{ch2lvg}%
        \includegraphics[width=.45\textwidth]{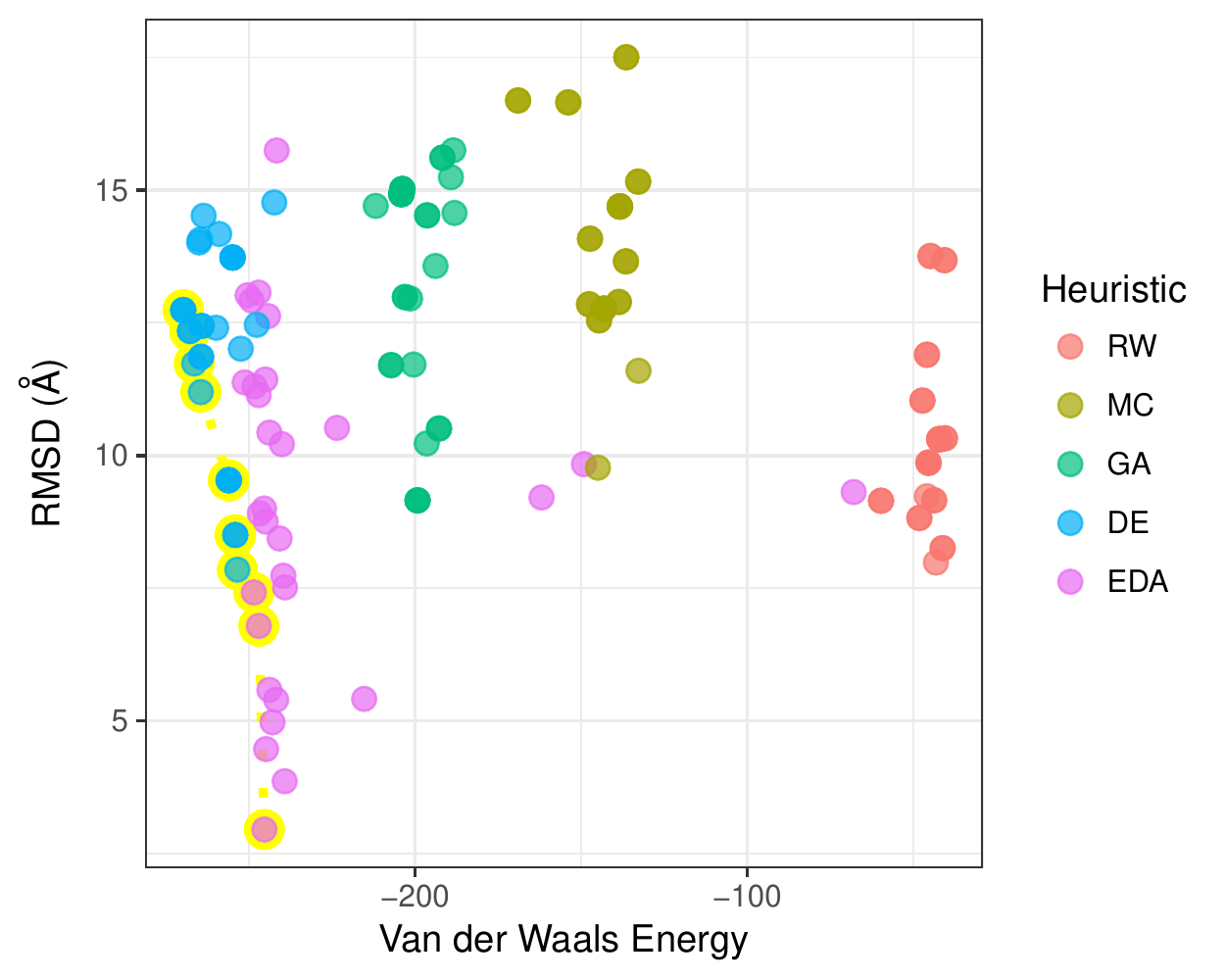}
 }\\
    \subfloat[][]{%
        \label{ch2kk7}%
        \includegraphics[width=.45\textwidth]{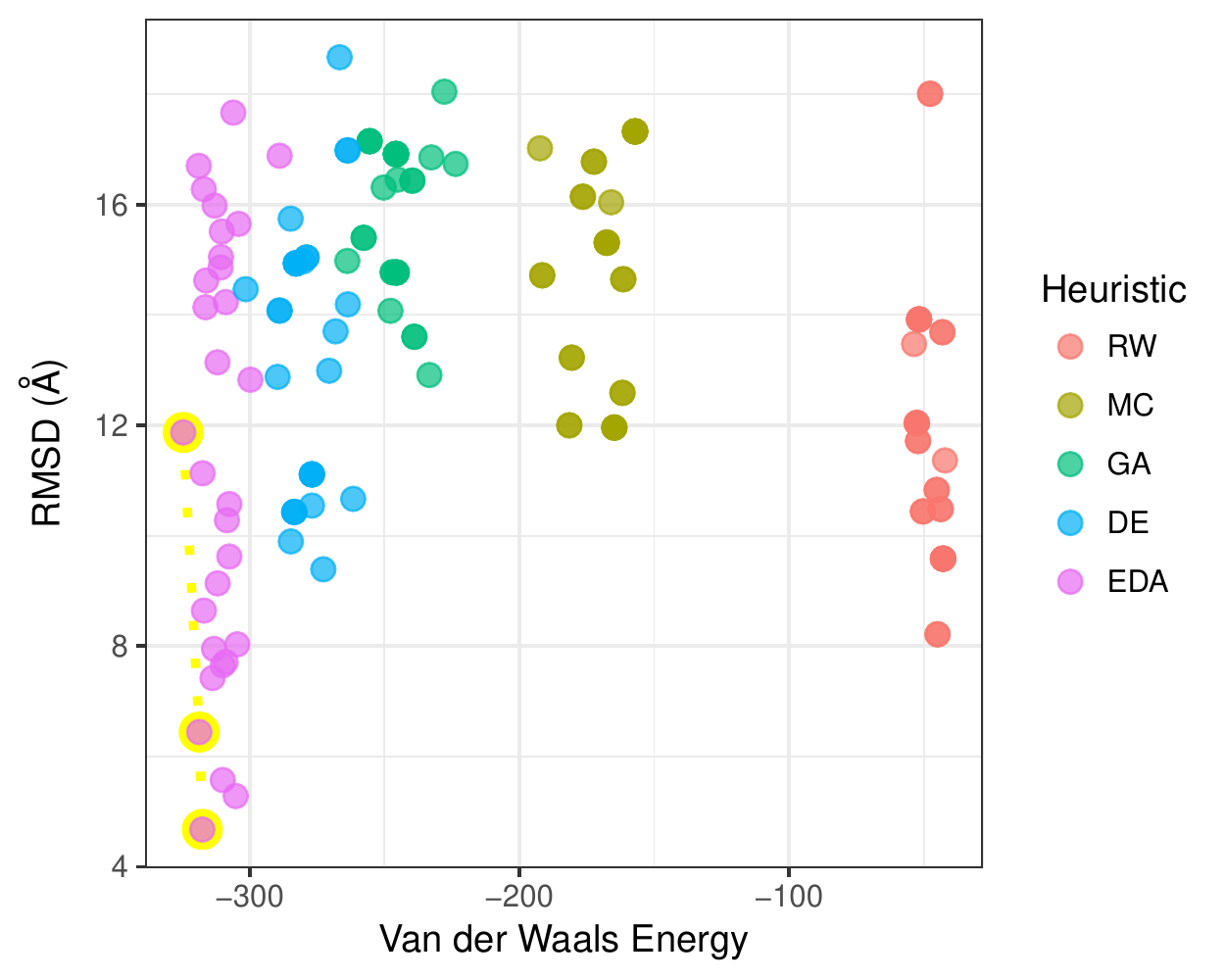}
}
    \subfloat[][]{%
        \label{ch2x43}%
        \includegraphics[width=.45\textwidth]{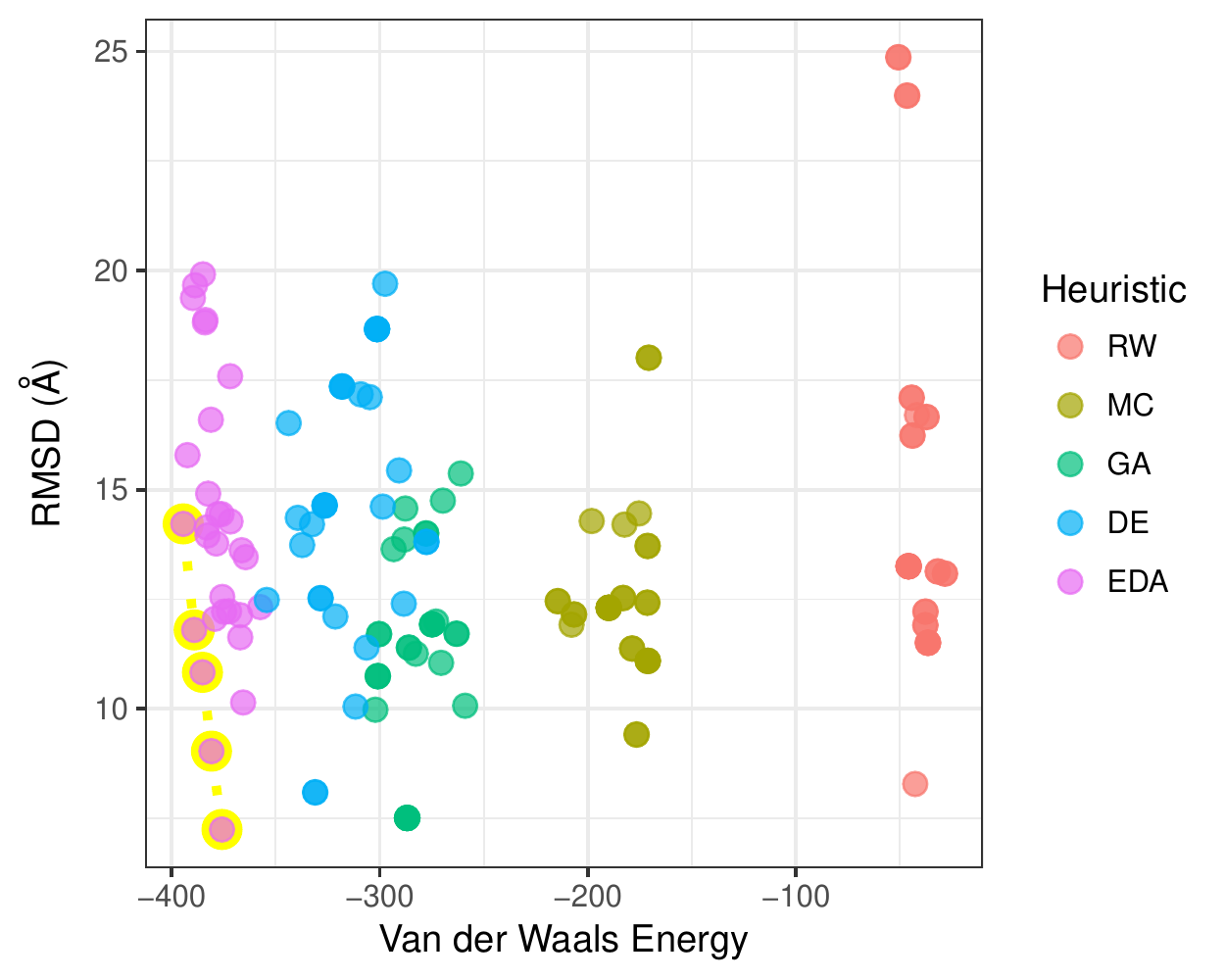}
 }

    \caption[EDA.]{Comparison of the energy values between our EDA with other heuristics:
    \subref{ch1a11} protein 1A11;
    \subref{ch2lvg} protein 2LVG;
    \subref{ch2kk7} protein 2KK7 and,
    \subref{ch2x43} protein 2X43.}%
    \label{ch}%
\end{figure*}

Figure~\ref{ch} shows the scatterplot between van der Waals energy and RMSD for all proteins used in this work.
As we expected, RW was the worst for all cases and MC was the second worst. Then, GA was worse than DE, although
GA got some points mixed with DE for the two larger proteins used (2KK7 and 2X43). Finally, the KDEO was better
for all cases considering only the RMSD.

For the smaller protein 1A11 (Figure~\ref{ch1a11}), DE was able to find better energy values than KDEO, although
the KDEO got a better RMSD. Then, for the protein 2LVG (Figure~\ref{ch2lvg}), DE also found better energy values.
However, the best RMSD found by DE was 7.847 and most of the RMSD values found by KDEO are below 7.775. For protein
2KK7 (Figure~\ref{ch2kk7}) the KDEO managed to find a better energy and RMSD than all other heuristics, while most
of the RMSD of other heuristics are higher than 8. Finally, for protein 2X43 (Figure~\ref{ch2x43}) the KDEO also
found better solutions with many points close to -400 energy and RMSD close to 10. No other heuristics managed to
appear in the Pareto Front. Therefore, it looks that the EDAs becomes more efficient as the protein size increases.

Each protein has its own set of parameters (population size, selection pressure etc) that works better, but in
these experiments, we fixed the same parameters for all four proteins, so that we can also evaluate how the
parameters would be for an unknown set of proteins without needing to calibrate them before.

\begin{figure*}%
    \centering
    \subfloat[][]{%
        \label{hg1a11}%
         \includegraphics[width=1\textwidth]{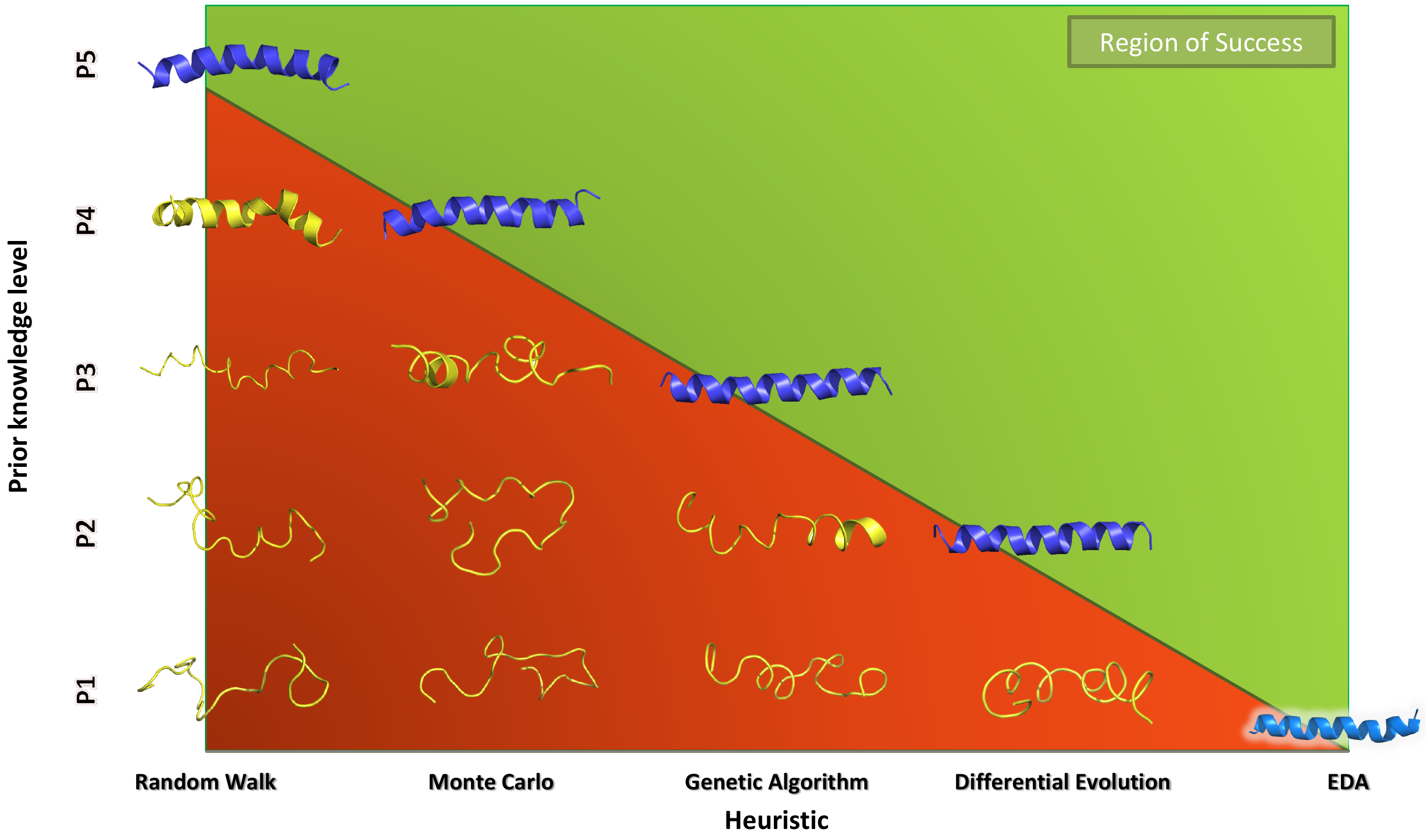}
 }
 \\
    \subfloat[][]{%
        \label{hr1a11}%
         \includegraphics[width=.5\textwidth]{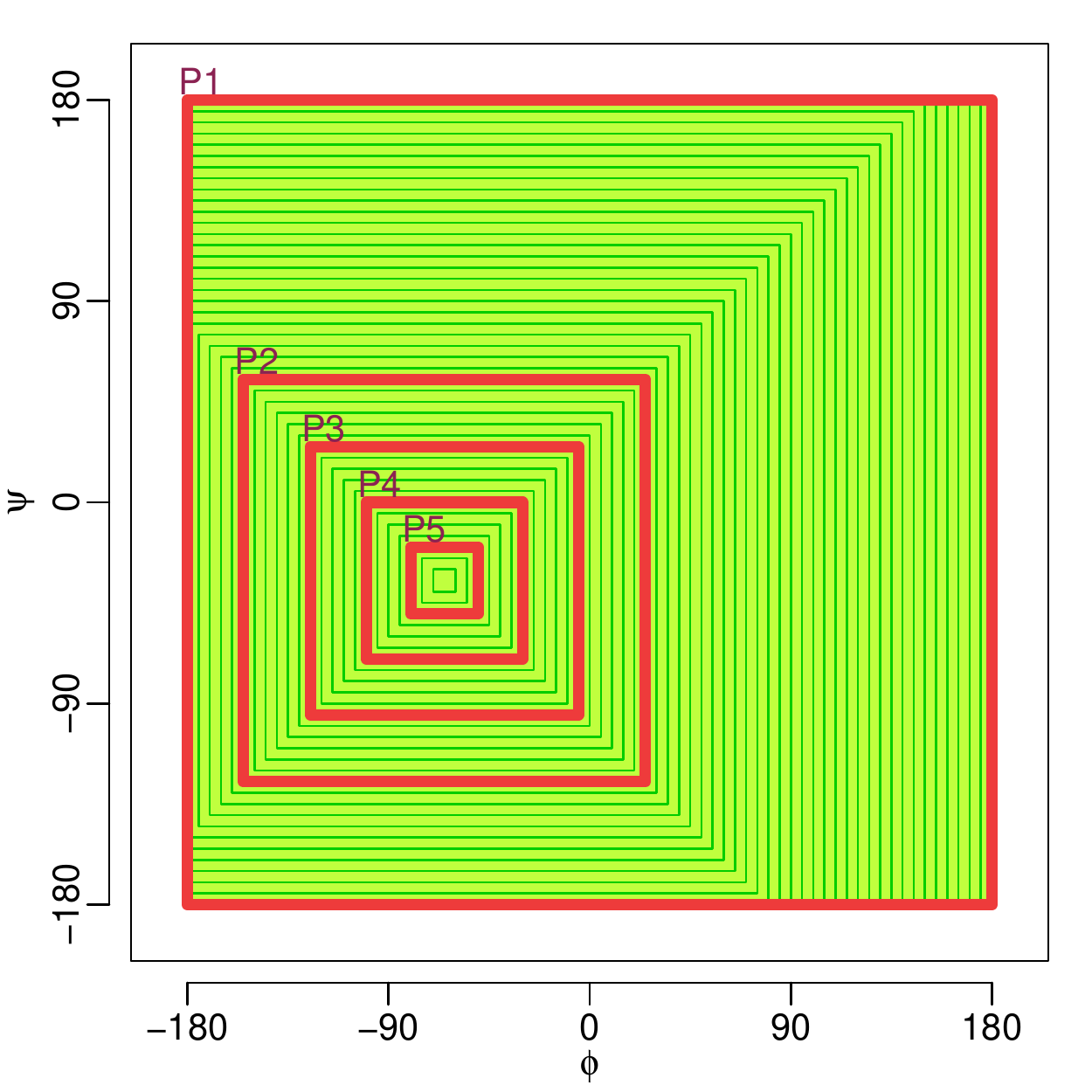}
 } \\

    \caption[EDA.]{Comparison of protein configurations between five different heuristics for protein 1A11:
    \subref{hg1a11} chart showing the success and the failure according to the prior-knowledge level and,
    \subref{hr1a11} the minimum search space needed for each heuristic to get success in their predictions.}%
    \label{h1a11}%
\end{figure*}

Finally, we ran another experiment in order to show the strength of the EDA against other heuristics. We created a hypothetical chart
that correlates the heuristic and the prior knowledge level needed to the heuristic be successful (Figure~\ref{hg1a11}). The vertical
axis is ordered by the prior knowledge level, i.e. how much the search space needs to be reduced so that the heuristic can hit the
correct solution. P1 means that the heuristic uses the whole search space when creating the initial population (as we have done in
all previous experiments, where it ranges from $(-180, +180)$ in $\phi$ and $\psi$ directions). Then we subsequently split the
search space using 5 degrees offset toward the right solution, in this case, the dihedral angles of protein 1A11 (look at Figure~\ref{ppDensity} and
see that the middle of the red lines is $[\phi, \psi] = [-65, -35]$). There were 49 divisions and all the levels are shown in Figure~\ref{hr1a11}.

Even a poor heuristic, as RW, managed to find the correct solution. However, the prior knowledge level needed by RW was so high that for any
input it will give almost the same output. We denote this prior knowledge level as P5, since RW would fail for any larger search space, that
is, this is the maximum search space we can use so that RW finds the correct solution. Using a slightly higher search space, MC managed to
find the correct solution. We call this level P4 and RW has failed at this level. Next, GA found the correct solution with level P3. At this
level, MC and RW have failed. In the next level, DE managed to find the right solution when other heuristics as GA, MC and RW have failed.
We denote this level as P2. Finally, the EDA (KDEO) using no prior knowledge (using the whole range of the search space) managed to find
the correct solution while all other heuristics have failed. That means the EDA was able to predict the correct protein configuration with no bias.

Furthermore, it is interesting to notice that the main difference between heuristics RW, MC, GA, DE and EDA (KDEO in this case) lies in the
amount of information they extract from their populations in order to sample new solutions. That is, RW does not use any information from its
population so the evolutionary process is completely random. MC uses information about one individual in order to sample new ones. GA uses
information about two individuals to compose the new ones, and DE uses information from three individuals. Thus, we have 0, 1, 2, and 3
individuals that are being used for RW, MC, GA and DE, respectively, to sample new solutions. We believe that this is related to the strength
of the heuristics as they appear in literature. That means, in general, DE works better than GA, which works better than MC, which works better than RW.

Therefore, as we know, the amount of information EDAs can extract from promising individuals in order to generate the offspring is much higher
than for DE. That happens because EDAs are designed to work with a set of selected individuals, which can have promising information to drive
the evolutionary process toward promising solutions. This is why we believe EDAs are suitable for any type of combinatory problem. Besides,
when dealing with problems where we already know how variables could behave, it becomes more effective for an EDA to find the correct solution.
This is what we have been doing with the PSP problem, associating the correlation between close variables together and creating probabilistic
models from a set of promising individuals.

\section{Conclusion}
\label{conclusions}

We develop three different probabilistic models for an Estimation of Distribution Algorithm specific for the \emph{ab initio} Protein Structure
Prediction with real-valued variables. We refer to these three methods as UNIO for the Univariate model-based Optimization, KDEO for the two-dimensional
Kernel Density Estimation model-based Optimization and FGMO for the Finite Gaussian Mixtures model-based Optimization, yielding three different EDAs
for PSP. The difference among these three EDAs is how they extract representative information from the selected individuals and use this information
to sample the offspring.

The first, UNIO, is the simplest one. It is similar to UMDA$_c$ but can handle problems with multiple modes instead of one, in an efficient manner.
Secondly, knowing that dihedral angles $\phi$ and $\psi$ of amino acids have a strong correlation we modeled the pair of dihedral angles $[\phi, \psi]$
as correlated for our two-dimensional probabilistic models: KDEO and FGMO. Results have shown that accounting for correlation enables us to find better
protein configurations in the search space.

In this work, only the van der Waals energy was used to compose the fitness function, so we can focus on the probabilistic model. Thus, we tried to use
only $\alpha$-helices in our predictions where van der Waals energy alone might represent the atomic interactions well. For the largest protein used in
the experiments, 2X43, the RMSD is still not as good as we expected. We believe that this may occur because van der Waals energy alone cannot handle
correctly the loop between the two helices. On the other hand, for the three smaller proteins 1A11, 2LVG and 2KK7 the configurations we found were
good enough, producing a low RMSD.

We also compared our EDA (more specifically, the KDEO) with other heuristics from literatures as Random Walk, Monte Carlo, Genetic Algorithm and
Differential Evolution. We noticed that the number of solutions used to compose new ones constitute a critical step in designing a good optimization
algorithm. For example, the weakest heuristic RW does not use any information about the evolutionary process to infer new solutions, so it is completely
random. The MC (usually better than RW) uses information about one single solution from the original population in order to infer new ones. The
GA (usually better than MC) makes use of two individuals from the population in order to compose the offspring. The DE (usually better than GA)
uses information about three individuals to infer the new ones. Thus, we noticed a trend in the number of individuals used to compose the offspring
and the strength of the optimization algorithm. This is why we believe EDAs are better and more efficient optimization techniques, since they use a
set of promising solutions (the selected individuals) and try to extract some representative statistical knowledge from this set (the probabilistic
model) in order to, finally, infer better solutions, leading the whole evolutionary process toward promising directions. We made this comparison with
the EDA against RW, MC, GA and DE to show why an EDA can perform better than these previous algorithms. Actually, we are not taking the RW as a serious
competitor against the EDA. We just wanted to emphasize the importance of using several potential solutions (the mechanism that the EDAs do) to build
the new ones, against other approaches that do not do this as illustrated in Figure~\ref{hg1a11}.

Until now, we had not seen an EDA for PSP with \emph{ab initio} and full-atom representation approaches. So we decided to use some knowledge of
the behavior of proteins, modeling the correlation $[\phi; \psi]$ and developing our specific EDA aimed at the PSP problem.

At the first glance, comparing the energy aspect between the two-dimensional probabilistic model-based techniques, KDEO and FGMO, it looks that
KDEO is better in most of the cases. However, we noticed that the number of mixture components used in FGMO has a high effect on the energy values.
Then, we still need to develop an automated and efficient way to estimate the ideal number of mixture components for each pair $[\phi; \psi]$.
Although FGMO needs an iterative algorithm in order to estimate the parameters (EM algorithm), the computational efficiency stays close to UNIO.
The main reason for this is because we are using the optimized PDF function (Equation~\ref{pdf2} instead of Equation~\ref{pdf1}) and a small set of selected.

Besides possible computational and statistical improvements, still the fitness function could be improved by adding other non-covalent energies as
solvation, hydrogen bonding and electrostatic energy to our experiments. However, before adding these energies one needs to ensure that such energies
are as efficient as possible, as already achieved for van der Waals and solvation energies in previous works \cite{Bonetti2014}. Otherwise, it would
be the bottleneck of the whole algorithm.

We also know that the energy effects acting in proteins are contradictory and sometimes operate on different scales. Thus, the next step is to
bring the Multi-Objective approach from the GA from \cite{Brasil2013} to our EDA.

%\vspace{20pt}
%{\bf Web-interface}

%There is a experimental web-interface for our proposed EDAs for PSP hosted at \\ \url{http://200.144.255.42:8004/}. On the left menu, it is
%necessary to choose the option \emph{ProtPred-EDA Tools}, selected the set of parameters (on the right panel) and then run the algorithms.

\appendix
\section{Statistical analysis}

We performed the pairwise Wilcoxon test in order to determine whether there is relevant difference between the evaluated methods.
The comparison was performed for every combination of methods for all the four evaluated proteins (1A11, 2LVG, 2LKK7 and 2X43).
The p-values that are greater than $0.05$ are highlighted in bold and they show that there is no significance difference between
them, according to the Wilcoxon test. Table~\ref{tabBest} shows the test for the van der Waals energy values. In this table,
there is a single value in bold, indicating that, according to the Wilcoxon test, most results concerning van der Waals energy
were different for all proteins evaluated. Table~\ref{tabRms} shows the test for the RMSD for the best protein configuration
found over all runs for each method. Although the Wilcoxon test for RW-KDEO and DE-KDEO got a high p-value for almost all
proteins, it is possible to see by looking at Figure~\ref{ch} that KDEO was the only capable method of finding RMSD values below $7.5$
for protein 2LVG (Figure~\ref{ch2lvg}) and below $8.0$ for protein 2KK7 (Figure~\ref{ch2kk7}). Finally, Table~\ref{tabRtime} shows the
comparison between p-values for the running times. For the protein 2X43, the running time between MC-KDEO produced no significant
difference; and for proteins 2LVG, 2KK7, 2X43 the running time between GA-KDEO also did not produce significant difference. That
means that KDEO can be used for PSP at a similar computational cost as a MC or a GA.

\input{img/tabBest.tex}

\input{img/tabRms.tex}

\input{img/tabRtime.tex}

\section*{Acknowledgment}

The authors would like to thank the Brazi\-lian research-funding a\-gen\-cy FA\-PESP for the fi\-nancial support given to this work.
Processes number: 2010\-/02928\--2 and 2013\--00919-4. Some part of this work was carried out while the first author visited
Durham University through the Durham/USP Research Exchange Programme.

\section*{References}

\bibliography{ref}
\end{document}

%% file: appendix/stat-tabFinalTable.tex
\begin{table}[h!t!]
\footnotesize
\caption{Best values out of 30 runs found for each protein and for each method. The columns Energy, RMSD and Time are not related to each other,  meaning that the values in a row may come from different running. The best values are highlighted.}
\label{tabFinalTable}
\begin{center}
\begin{tabular}{c|c|c|c|c|c}

 Protein & Method & Energy (kcal/mol) & RMSD (\AA) & Time (min.) \\ \hline \hline
   1a11 & RW &  -40.4 &  5.160 & 	 24.6 \\
   1a11 & MC & -113.3 &  6.335 & 	 36.0 \\
   1a11 & GA & -144.7 &  4.709 & 	 39.0 \\
   1a11 & DE & -167.9 &  2.976 & 	 37.1 \\
  1a11 & UNIO & -148.9 &  3.677 & 	 \bf{24.4} \\
1a11 & KDEO & -156.9 &  2.464 & 	 42.6 \\
  1a11 & FGMO & \bf{-162.5} &  \bf{2.338} & 	 24.6 \\ \hline
   2lvg & RW &  -59.7 &  7.981 & 	 52.8 \\
   2lvg & MC & -169.0 &  9.768 & 	 55.5 \\
   2lvg & GA & -211.8 &  9.153 & 	 61.1 \\
   2lvg & DE & -270.7 &  7.847 & 	 47.3 \\
  2lvg & UNIO & -233.0 &  5.605 & 	 \bf{34.4} \\
2lvg & KDEO & -255.4 &  \bf{2.953} & 	 79.1 \\
  2lvg & FGMO & \bf{-268.8} &  5.593 & 	 42.7 \\ \hline
   2kk7 & RW &  -53.6 &  8.213 & 	 68.6 \\
   2kk7 & MC & -192.3 & 11.959 & 	 64.7 \\
   2kk7 & GA & -263.7 & 12.911 & 	 85.6 \\
   2kk7 & DE & -301.4 &  9.391 & 	 49.0 \\
  2kk7 & UNIO & -275.8 &  6.655 & 	 \bf{43.3} \\
2kk7 & KDEO & -324.6 &  \bf{4.676} & 	 78.2 \\
  2kk7 & FGMO & \bf{-345.3} &  5.995 & 	 49.9 \\ \hline
   2x43 & RW &  -50.7 &  8.283 & 	 92.0 \\
   2x43 & MC & -214.4 &  9.412 & 	 90.0 \\
   2x43 & GA & -302.0 &  7.507 & 	104.4 \\
   2x43 & DE & -354.2 &  8.092 & 	 62.4 \\
  2x43 & UNIO & -325.5 &  9.257 & 	 \bf{56.3} \\
2x43 & KDEO & -394.4 &  \bf{7.245} & 	128.3 \\
  2x43 & FGMO & \bf{-413.1} &  9.710 & 	 82.8 \\
\end{tabular}
\end{center}
\end{table}

%% file: img/tabBest.tex
\begin{table}[h!t!]
\footnotesize
\caption[P-value comparison with pairwise Wilcoxon test evaluating the van der Waals energy.]{P-value comparison with pairwise Wilcoxon test evaluating the van der Waals energy}
\label{tabBest}
\begin{center}
\begin{tabular}{c|c|c|c|c}

 & 1A11 & 2LVG & 2KK7 & 2X43 \\ \hline \hline
    RW-MC  &        6.3e-10 &       5.9e-10 &       5.8e-10 &       5.9e-10 \\
    RW-GA  &        6.3e-10 &       5.9e-10 &       5.8e-10 &       5.9e-10 \\
    RW-DE  &        6.3e-10 &       5.9e-10 &       5.8e-10 &       5.9e-10 \\
%  RW-UNIO  &        6.3e-10 &       1.3e-09 &       5.8e-10 &       5.9e-10 \\
  RW-KDEO  &        6.3e-10 &       5.9e-10 &       5.8e-10 &       5.9e-10 \\
%  RW-FGMO  &        6.3e-10 &       5.9e-10 &       5.8e-10 &       4.9e-09 \\
    MC-GA  &        6.3e-10 &       5.9e-10 &       5.8e-10 &       5.9e-10 \\
    MC-DE  &        6.3e-10 &       5.9e-10 &       5.8e-10 &       5.9e-10 \\
 % MC-UNIO  &        6.3e-10 &       1.3e-07 &       5.8e-10 &       5.9e-10 \\
  MC-KDEO  &        6.3e-10 &       9.5e-09 &       5.8e-10 &       5.9e-10 \\
%  MC-FGMO  &        6.3e-10 & \bf{8.9e-01} & \bf{4.5e-01} &       1.2e-03 \\
    GA-DE  &        6.3e-10 &       5.9e-10 &       5.8e-10 &       1.6e-07 \\
%  GA-UNIO  &        4.3e-09 &       1.7e-04 &       5.6e-06 &       3.0e-07 \\
  GA-KDEO  &        6.3e-10 &       7.3e-07 &       5.8e-10 &       5.9e-10 \\
%  GA-FGMO  &        1.0e-08 & \bf{1.4e-01} &       1.8e-03 &       9.1e-06 \\
%  DE-UNIO  &        6.3e-10 &       5.9e-10 &       2.9e-09 & \bf{9.6e-02} \\
  DE-KDEO  &        4.3e-09 &       3.1e-09 &       5.8e-10 &       5.9e-10 \\
%  DE-FGMO  &        9.7e-07 &       1.8e-06 &       2.7e-04 &       1.5e-06 \\
UNIO-KDEO  &        2.8e-07 &       2.7e-07 &       5.8e-10 &       5.9e-10 \\
UNIO-FGMO  &        3.2e-06 & \bf{2.0e-01} &       3.7e-04 &       1.5e-06 \\
KDEO-FGMO  &        4.1e-02 &       3.8e-02 &       1.3e-04 &       4.2e-07 \\
\end{tabular}
\end{center}
\end{table}

%% file: img/tabRms.tex
\begin{table}[h!t!]
\footnotesize
\caption[P-value comparison with pairwise Wilcoxon test evaluating the RMSD.]{P-value comparison with pairwise Wilcoxon test evaluating the RMSD}
\label{tabRms}
\begin{center}
\begin{tabular}{c|c|c|c|c}

 & 1A11 & 2LVG & 2KK7 & 2X43 \\ \hline \hline
    RW-MC  &        7.3e-04 &       5.4e-07 &       2.1e-04 &       1.4e-01 \\
    RW-GA  &  \bf{1.0e+00} &       5.8e-05 &       7.2e-07 &       2.2e-03 \\
    RW-DE  &  \bf{1.0e+00} &       1.9e-02 & \bf{9.4e-02} &       1.0e+00 \\
%  RW-UNIO  &  \bf{1.0e+00} & \bf{1.0e+00} & \bf{4.7e-01} & \bf{1.0e+00} \\
  RW-KDEO  &        1.8e-02 & \bf{5.9e-01} & \bf{1.0e+00} & \bf{1.0e+00} \\
%  RW-FGMO  &  \bf{1.6e-01} & \bf{1.0e+00} & \bf{1.0e+00} &       1.8e-02 \\
    MC-GA  &        9.3e-03 & \bf{1.0e+00} & \bf{8.6e-01} & \bf{1.4e-01} \\
    MC-DE  &        3.4e-03 &       6.0e-04 & \bf{9.5e-02} &       7.8e-03 \\
%  MC-UNIO  &        2.3e-04 &       4.6e-04 &       1.6e-02 & \bf{1.9e-01} \\
  MC-KDEO  &        1.3e-04 &       4.9e-07 &       1.2e-02 & \bf{2.8e-01} \\
%  MC-FGMO  &  \bf{2.4e-01} &       2.1e-03 &       3.6e-02 &       3.4e-06 \\
    GA-DE  &  \bf{1.0e+00} &       4.2e-02 &       3.0e-03 &       2.2e-04 \\
%  GA-UNIO  &  \bf{5.1e-01} &       3.3e-03 &       4.8e-06 &       9.4e-03 \\
  GA-KDEO  &        8.4e-03 &       1.7e-05 &       1.8e-04 &       4.5e-03 \\
%  GA-FGMO  &  \bf{1.0e+00} &       2.1e-02 &       1.2e-04 &       2.7e-07 \\
%  DE-UNIO  &  \bf{1.0e+00} & \bf{6.3e-01} & \bf{1.0e+00} & \bf{1.0e+00} \\
  DE-KDEO  &  \bf{8.0e-02} &       1.5e-03 & \bf{8.6e-01} & \bf{1.0e+00} \\
%  DE-FGMO  &  \bf{1.7e-01} & \bf{1.0e+00} & \bf{1.0e+00} & \bf{6.7e-02} \\
UNIO-KDEO  &        2.9e-02 & \bf{2.2e-01} & \bf{1.0e+00} & \bf{1.0e+00} \\
UNIO-FGMO  &        4.0e-02 & \bf{1.0e+00} & \bf{1.0e+00} &       3.2e-03 \\
KDEO-FGMO  &        3.3e-03 & \bf{5.2e-02} & \bf{1.0e+00} &       9.2e-03 \\
\end{tabular}
\end{center}
\end{table}

%% file: img/tabRtime.tex
\begin{table}[h!t!]
\footnotesize
\caption[P-value comparison with pairwise Wilcoxon test evaluating the running time.]{P-value comparison with pairwise Wilcoxon test evaluating the running time}
\label{tabRtime}
\begin{center}
\begin{tabular}{c|c|c|c|c}

 & 1A11 & 2LVG & 2KK7 & 2X43 \\ \hline \hline
    RW-MC  &        3.6e-16 &       3.6e-16 &       2.9e-13 &       1.1e-15 \\
    RW-GA  &        3.6e-16 &       1.2e-11 &       2.1e-09 &       2.0e-08 \\
    RW-DE  &        3.6e-16 & \bf{8.8e-01} & \bf{7.9e-01} &       1.0e+00 \\
%  RW-UNIO  &        3.6e-16 &       2.1e-07 & \bf{9.6e-02} & \bf{1.0e+00} \\
  RW-KDEO  &        3.6e-16 &       1.4e-13 &       1.8e-06 &       6.8e-16 \\
%  RW-FGMO  &        3.6e-16 & \bf{1.0e+00} &       3.5e-05 & \bf{1.0e+00} \\
    MC-GA  &  \bf{6.3e-01} &       5.8e-04 & \bf{7.1e-01} & \bf{1.0e+00} \\
    MC-DE  &  \bf{6.2e-01} &       1.4e-07 &       1.0e-09 &       9.0e-14 \\
%  MC-UNIO  &  \bf{1.0e+00} &       4.0e-12 &       4.5e-14 &       6.8e-16 \\
  MC-KDEO  &        8.8e-14 &       4.4e-02 &       5.4e-05 & \bf{2.8e-01} \\
%  MC-FGMO  &        8.3e-06 &       6.4e-14 &       4.3e-15 &       1.1e-15 \\
    GA-DE  &  \bf{1.0e+00} &       4.4e-02 &       2.9e-07 &       2.4e-08 \\
%  GA-UNIO  &  \bf{9.7e-01} &       3.5e-02 &       1.8e-10 &       1.5e-10 \\
  GA-KDEO  &        2.7e-13 & \bf{2.4e-01} & \bf{5.8e-02} & \bf{1.0e+00} \\
%  GA-FGMO  &        4.8e-07 &       1.6e-09 &       1.1e-12 &       1.3e-10 \\
%  DE-UNIO  &  \bf{9.7e-01} & \bf{1.0e+00} & \bf{7.9e-01} & \bf{1.0e+00} \\
  DE-KDEO  &        6.6e-12 &       1.6e-03 &       9.1e-04 &       6.8e-16 \\
%  DE-FGMO  &        3.4e-07 & \bf{3.0e-01} & \bf{5.8e-02} & \bf{1.0e+00} \\
UNIO-KDEO  &        1.5e-12 &       1.4e-11 &       1.8e-07 &       3.6e-16 \\
UNIO-FGMO  &        1.7e-06 &       1.9e-05 & \bf{9.6e-02} & \bf{1.0e+00} \\
KDEO-FGMO  &        1.0e-15 &       8.3e-12 &       3.5e-10 &       6.8e-16 \\
\end{tabular}
\end{center}
\end{table}